\PassOptionsToPackage{unicode}{hyperref}
\PassOptionsToPackage{hyphens}{url}
\PassOptionsToPackage{dvipsnames,svgnames,x11names}{xcolor}
\documentclass[
  12pt]{article}

\usepackage{xparse}

\date{}

\newcommand{\secref}[1]{\hyperref[#1]{Section~\ref*{#1}}}
\newcommand{\figref}[1]{\hyperref[#1]{Figure~\ref*{#1}}}
\usepackage{tikz-cd} 
\usetikzlibrary{babel} 
\usetikzlibrary{shapes.geometric, arrows}  
\usetikzlibrary{arrows.meta,
                chains,
                positioning,
                fit,
                shapes.geometric,
                angles,quotes,
                decorations.pathreplacing
                }
\usepackage{listings}
\usepackage{amsmath,amsthm,amssymb,amsfonts,mathrsfs,graphicx,color,bbm,bm}
\usepackage{enumitem} 
\usepackage{iftex}
\ifPDFTeX
  \usepackage[T1]{fontenc}
  \usepackage[utf8]{inputenc}
  \usepackage{textcomp} 
\else 
  \usepackage{unicode-math}
  \defaultfontfeatures{Scale=MatchLowercase}
  \defaultfontfeatures[\rmfamily]{Ligatures=TeX,Scale=1}
\fi
\usepackage{lmodern}
\ifPDFTeX\else  
\fi
\IfFileExists{upquote.sty}{\usepackage{upquote}}{}
\IfFileExists{microtype.sty}{
  \usepackage[]{microtype}
  \UseMicrotypeSet[protrusion]{basicmath} 
}{}
\makeatletter
\@ifundefined{KOMAClassName}{
  \IfFileExists{parskip.sty}{%
    \usepackage{parskip}
  }{
    \setlength{\parindent}{0pt}
    \setlength{\parskip}{6pt plus 2pt minus 1pt}}
}{
  \KOMAoptions{parskip=half}}
\makeatother
\usepackage{xcolor}
\makeatletter
\ifx\paragraph\undefined\else
  \let\oldparagraph\paragraph
  \renewcommand{\paragraph}{
    \@ifstar
      \xxxParagraphStar
      \xxxParagraphNoStar
  }
  \newcommand{\xxxParagraphStar}[1]{\oldparagraph*{#1}\mbox{}}
  \newcommand{\xxxParagraphNoStar}[1]{\oldparagraph{#1}\mbox{}}
\fi
\ifx\subparagraph\undefined\else
  \let\oldsubparagraph\subparagraph
  \renewcommand{\subparagraph}{
    \@ifstar
      \xxxSubParagraphStar
      \xxxSubParagraphNoStar
  }
  \newcommand{\xxxSubParagraphStar}[1]{\oldsubparagraph*{#1}\mbox{}}
  \newcommand{\xxxSubParagraphNoStar}[1]{\oldsubparagraph{#1}\mbox{}}
\fi
\makeatother

\usepackage{longtable,booktabs,array}
\usepackage{calc} 
\usepackage{etoolbox}
\makeatletter
\patchcmd\longtable{\par}{\if@noskipsec\mbox{}\fi\par}{}{}
\makeatother
\IfFileExists{footnotehyper.sty}{\usepackage{footnotehyper}}{\usepackage{footnote}}
\makesavenoteenv{longtable}
\usepackage{graphicx}
\makeatletter
\def\maxwidth{\ifdim\Gin@nat@width>\linewidth\linewidth\else\Gin@nat@width\fi}
\def\maxheight{\ifdim\Gin@nat@height>\textheight\textheight\else\Gin@nat@height\fi}
\makeatother
\setkeys{Gin}{width=\maxwidth,height=\maxheight,keepaspectratio}
\makeatletter
\def\fps@figure{htbp}
\makeatother

\makeatletter
\@ifpackageloaded{caption}{}{\usepackage{caption}}
\AtBeginDocument{%
\ifdefined\contentsname
  \renewcommand*\contentsname{Table of contents}
\else
  \newcommand\contentsname{Table of contents}
\fi
\ifdefined\listfigurename
  \renewcommand*\listfigurename{List of Figures}
\else
  \newcommand\listfigurename{List of Figures}
\fi
\ifdefined\listtablename
  \renewcommand*\listtablename{List of Tables}
\else
  \newcommand\listtablename{List of Tables}
\fi
\ifdefined\figurename
  \renewcommand*\figurename{Figure}
\else
  \newcommand\figurename{Figure}
\fi
\ifdefined\tablename
  \renewcommand*\tablename{Table}
\else
  \newcommand\tablename{Table}
\fi
}
\@ifpackageloaded{float}{}{\usepackage{float}}
\floatstyle{ruled}
\@ifundefined{c@chapter}{\newfloat{codelisting}{h}{lop}}{\newfloat{codelisting}{h}{lop}[chapter]}
\floatname{codelisting}{Listing}

\makeatother
\makeatletter
\@ifpackageloaded{caption}{}{\usepackage{caption}}
\@ifpackageloaded{subcaption}{}{\usepackage{subcaption}}
\makeatother

\ifLuaTeX
  \usepackage{selnolig}  
\fi
\usepackage[]{natbib}
\usepackage{bookmark}

\IfFileExists{xurl.sty}{\usepackage{xurl}}{} 
\hypersetup{
  pdftitle={Title},
  pdfauthor={Author 1; Author 2},
  pdfkeywords={3 to 6 keywords, that do not appear in the title},
  colorlinks=true,
  linkcolor={blue},
  filecolor={Maroon},
  citecolor={Blue},
  urlcolor={Blue},
  pdfcreator={LaTeX via pandoc}}
\hypersetup{hidelinks}

\newcommand{\anon}{1}

\usepackage{authblk}

\DeclareCaptionFormat{listing}{\centering Code Block \arabic{lstlisting}: #3}
\NewDocumentCommand{\coderef}{m o}{%
  \hyperref[#1]{%
    \IfNoValueTF{#2}{%
      Code Block~\ref*{#1}%
    }{%
      line~#2%
    }%
  }%
}

\newcommand{\textandreference}[2]{\texorpdfstring{\hyperref[#2]{#1\ref*{#2}}}{#1\ref*{#2}}}

\newcommand{\lbfig}[1]{\label{fig:#1}}
\newcommand{\reffig}[1]{\textandreference{Figure~}{fig:#1}}

\newcommand{\spacednabla}[2]{\nabla_{\!#1}\,{#2}}

\renewcommand{\P}{\mathbb{P}}

\renewcommand{\phi}{\varphi}

\let\epsilon\varepsilon

\newcommand{\blank}[1]{}

\newcommand{\Poisson}{\mathrm{Poisson}}

\usepackage[ruled, lined, longend, linesnumbered]{algorithm2e}

\makeatletter \def\algocf@capseparator{}\makeatother

\begin{document}

\def\spacingset#1{\renewcommand{\baselinestretch}%
{#1}\small\normalsize} \spacingset{1}

\if1\anon
{
  \title{\bf A general framework for computation and estimation using the saddlepoint approximation}
  \author[1]{Godrick Oketch}
  \author[2]{Rachel M. Fewster}
  \author[2]{Jesse Goodman\thanks{Email: jesse.goodman@auckland.ac.nz
  \newline The authors gratefully acknowledge funding support from the Royal Society of New Zealand Marsden fund.}}
  \affil[1]{School of Public Health, University of California, Berkeley, California, U.S.A.}
  \affil[2]{Department of Statistics, University of Auckland, Private Bag 92019, Auckland, New Zealand}
  \maketitle
} \fi

\if0\anon
{
  \bigskip
  \bigskip
  \bigskip
  \begin{center}
    {\LARGE\bf A general framework for computation and estimation using the saddlepoint approximation}
\end{center}
  \medskip
} \fi

\bigskip
\begin{abstract}
The saddlepoint approximation provides highly accurate approximations to probability density and mass functions using only the corresponding moment generating functions (MGFs). Recent work has increasingly seen the saddlepoint approximation applied to likelihood functions, enabling likelihood-based inference in models where exact likelihoods are intractable. However, existing implementations have largely been developed on a model-by-model basis, and the methodology remains underutilized because of the conceptual and computational challenges of working with MGFs. 
We introduce a unified framework for model construction and computation using the saddlepoint approximation. The framework is based on a collection of model-building operations that preserve access to MGFs while allowing complex distributions to be constructed from simpler components.  With these components, users need only provide a high-level specification of the model structure, from which the software automatically assembles the necessary generating functions, saddlepoints, and gradients, and performs the optimization of the saddlepoint likelihood. 
We also introduce a diagnostic that quantifies the difference between saddlepoint and exact likelihood estimates, even when the exact likelihood is
unavailable. 
The framework is implemented in the R package \texttt{saddlepoint} and provides fast, convenient computation of parameter estimates, standard errors, and the discrepancy diagnostic. Numerous examples illustrate the scope and flexibility of the approach.
\end{abstract}

\noindent%
{\it Keywords:} 
approximate likelihood, 
constrained optimization,
cumulant generating function,
latent identity model,
non-invertible transformation, 
statistical linear inverse model 
\vfill

\newpage
\spacingset{1.8} 

\section{Introduction}\label{sec:intro}

The saddlepoint approximation is a method for computing an approximate probability density or mass function of a random variable, given only its moment generating function (MGF). 
Recently, there has been considerable interest in using the saddlepoint approximation as an estimation tool, by treating the saddlepoint density as an approximation to the model likelihood. Saddlepoint-based estimation has been shown to be an accurate and computationally efficient estimation option for a diverse range of models in which exact likelihood computations are intractable, but the associated MGF is available.

Let $\boldsymbol{Y}$ be a $d$-dimensional random variable that depends on a parameter vector $\boldsymbol{\theta}$ and has probability density or mass function $f(\boldsymbol{y};\boldsymbol{\theta})$. Define the MGF and cumulant generating function (CGF) of $\boldsymbol{Y}$ by
\[
     M_Y(\boldsymbol{t};\boldsymbol{\theta}) = \mathbb{E}(e^{\boldsymbol{tY}}) ~ \text{and}~  K_Y(\boldsymbol{t};\boldsymbol{\theta}) = \log(M_Y(\boldsymbol{t};\boldsymbol{\theta})),
\]
respectively, 
such that $\boldsymbol{t}\in\mathbb{R}^{d}$ belongs to an open interval containing zero in which the MGF converges.
By convention, $\boldsymbol{t}$ is interpreted as a row vector and $\boldsymbol{Y}$ as a column vector.
For $\boldsymbol{y}\in\mathbb{R}^{d}$, \citet{Daniels1954}
gave the \textit{saddlepoint approximation} to $f(\boldsymbol{y}; \boldsymbol{\theta})$ as
\begin{equation}\label{eq:Saddlepoint_Defn}
    \hat{f}(\boldsymbol{y};\boldsymbol{\theta}) = \frac{\exp(K_Y(\hat{\boldsymbol{t}};\boldsymbol{\theta})-\hat{\boldsymbol{t}}\boldsymbol{y})}
{\sqrt{\det(2\pi K_Y''(\hat{\boldsymbol{t}};\boldsymbol{\theta}) )}},
\end{equation}
where $\hat{\boldsymbol{t}} = \hat{\boldsymbol{t}}(\boldsymbol{\theta};\boldsymbol{y})$ is the solution of the \textit{saddlepoint equation} defined by
\begin{equation}\label{eqn:saddlepoint_eqn}
    K_Y'(\hat{\boldsymbol{t}};\boldsymbol{\theta}) = \boldsymbol{y}
    .
\end{equation}
The notations $K'_Y$ and $K''_Y$ represent the first and second-order gradients of the CGF with respect to $\boldsymbol{t}$. 

To obtain estimates of $\boldsymbol{\theta}$ given observations of $\boldsymbol{y}$, \eqref{eq:Saddlepoint_Defn} is interpreted as a likelihood function:
\begin{equation}\label{eq:Likelihood_Fn}
    \hat{L}(\boldsymbol{\theta}\,|\, \boldsymbol{y}) = \hat{f}(\boldsymbol{y};\boldsymbol{\theta})=
    \frac{\exp(K_Y(\hat{\boldsymbol{t}};\boldsymbol{\theta})-\hat{\boldsymbol{t}}\boldsymbol{y})}
{\sqrt{\det(2\pi K_Y''(\hat{\boldsymbol{t}};\boldsymbol{\theta}) )}}
.
\end{equation}
We refer to $\hat{L}(\boldsymbol{\theta}\,|\, \boldsymbol{y})$ as the saddlepoint approximation to the likelihood function, or simply the \emph{saddlepoint likelihood}. 
Maximizing $\hat{L}(\boldsymbol{\theta}\,|\, \boldsymbol{y})$ yields an approximation to the maximum likelihood estimate, which we call the \emph{saddlepoint MLE}.
Collectively, we refer to the steps of finding the CGF, forming the saddlepoint likelihood, and maximizing it to obtain an estimate of $\boldsymbol{\theta}$ as the \emph{saddlepoint likelihood method}.

Recent applications of the saddlepoint likelihood method include integer-valued autoregressive models \citep{Pedeli2015}, abundance estimation \citep{Zhang2019, Zhang2021, Zhang2022}, and population growth models \citep{Davison2020}. 
Further, \cite{Goodman2022} demonstrated that under appropriate hypotheses---where $n$ relates to a specific limiting framework scale---estimates from the saddlepoint likelihood 
differ from those of the true likelihood by a minimal order of 
$n^{-2}$, $n^{-3/2}$, or $n^{-1}$, depending on the model characteristics.
These small discrepancies, when compared to the scale of inferential uncertainty, strongly support the effectiveness of \eqref{eq:Likelihood_Fn}.

The key difference in the shift from a saddlepoint approximation \eqref{eq:Saddlepoint_Defn} to the saddlepoint likelihood \eqref{eq:Likelihood_Fn} is that $\boldsymbol{\theta}$ is no longer fixed. 
Consequently, the formulation of the CGF $K_Y(\hat{\boldsymbol{t}};\boldsymbol{\theta})$ and its gradients $K'_Y(\hat{\boldsymbol{t}};\boldsymbol{\theta})$ and $K''_Y(\hat{\boldsymbol{t}};\boldsymbol{\theta})$ must also consider $\boldsymbol{\theta}$ as a variable, recognizing that the CGF depends on $\boldsymbol{\theta}$ both directly and indirectly through $\hat{\boldsymbol{t}}$.

In practice, computation of the CGF for an observable 
variable $\boldsymbol{Y}$ is not always straightforward; it depends on the model structure and can sometimes involve several other CGFs.
Such challenges extend to the first and second-order gradients of the CGF. 
Additionally, the implicit function of $\boldsymbol{\theta}$ via \eqref{eqn:saddlepoint_eqn}complicates the optimization of the saddlepoint likelihood. In some models, further complexities arise due to constraints on the elements of $\boldsymbol{t}$ and $\boldsymbol{\theta}$.

Our aim is to simplify these computations, allowing modelers to perform saddlepoint-based estimation using familiar tools such as R \citep{R2022} without needing to engage with the computational complexities of saddlepoint methods. 
We introduce an extensible computational framework that automates the computation of CGFs, saddlepoint likelihoods, and saddlepoint MLEs.
The framework provides CGF-building tools that directly match the structural components of a model, so modelers only need to specify the model structure after which the software works in the background to automatically assemble the CGF and its gradients. 
Once the CGF is assembled, automatic differentiation is used for accurate gradient-based optimization with respect to $\boldsymbol{\theta}$. We adopt a constrained optimization strategy to accommodate any model-specific constraints on $\boldsymbol{t}$ or $\boldsymbol{\theta}$. 
Overall, our framework enables practitioners to implement the saddlepoint likelihood method in just a few lines of code.

\section{Model building blocks for saddlepoint likelihoods}\label{sec:ModelBuildingBlocks}

In this section we identify model structures for which the saddlepoint approximation is often useful, and illustrate them with examples from recent literature. The common characteristic of these models is that the observed data variable $\boldsymbol{Y}$ is expressed as
a function of one or more unobservable random variables, denoted by $\boldsymbol{X}$. 
Often $\boldsymbol{X}$ arises naturally in the modeling process and has a simple and explicit distributional form, but what is observed as $\boldsymbol{Y}$ is some non-invertible operation applied to $\boldsymbol{X}$, e.g., a thinning operation or linear mapping to a lower-dimensional space. Thus, the variable that can be naturally modeled is unobservable, whereas the observable variable cannot be naturally modeled.

It is often difficult or impossible to compute the exact likelihood of $\boldsymbol{Y}$, but if the CGF of $\boldsymbol{Y}$ is computable we can use the saddlepoint likelihood instead. In this section we outline a variety of operations for which the CGF of $\boldsymbol{Y}$ is generally accessible and the saddlepoint likelihood method can be applied. We refer to these scenarios as model building blocks. These building blocks will subsequently be used to guide the design of our computational framework.

\subsection{Sum of independent and identically distributed variables}\label{sec:sum-of-iid}

Consider an observable random variable $\boldsymbol{Y}$ as the sum of $n$ independent and identically distributed (i.i.d.)\ copies of a 
$d$-dimensional
random variable $\boldsymbol{X}$, i.e., 
\begin{equation}\label{eq:sum_of_iid_rvs}
    \boldsymbol{Y} = \boldsymbol{X}_1 + \dotsb + \boldsymbol{X}_n.
\end{equation}
Observing $\boldsymbol{Y}$ alone is not enough to recover the summands $\boldsymbol{X}_i$. However, if the $\boldsymbol{X}_i$ have a known parametric distribution, the distribution of $\boldsymbol{Y}$ is determined implicitly by \eqref{eq:sum_of_iid_rvs}. It may not be straightforward to compute the likelihood function for $\boldsymbol{Y}$, but we can estimate the unknown parameters using the saddlepoint likelihood method because
the CGF of $\boldsymbol{Y}$ is attainable from the known CGF of $\boldsymbol{X}$ and the summation operation: $K_Y(\boldsymbol{t}) = n K_X(\boldsymbol{t})$. 

This application of the saddlepoint method was used by \citet{Davison2020} for parameter estimation in a linear birth-death branching process.
The motivation for their work was that the exact likelihood function for this branching process, which is known and has an explicit form, can be numerically unstable. 
The variable $\boldsymbol{Y}$ in their setup is the population size at the next step, given that at the current step, which by the branching property is the sum of i.i.d.\ copies of the offspring random variable, similar to \eqref{eq:sum_of_iid_rvs}.
The authors used the saddlepoint likelihood method to uncover the parameters of the offspring distribution.

\subsection{Thinning and splitting}\label{sec:Thinning}
Thinning and splitting are two related concepts that involve sub-sampling and partitioning of random variables, respectively.
Binomial thinning, or simply thinning, begins with a scalar 
variable $X$ that counts items of interest, of which only a fraction $p$ are observable.
The count of observed items, denoted $Y$, satisfies $(Y\,|\,X) \sim \mathrm{Binomial}(X, p)$. The distribution of $Y$ is called the $p$-thinning of the distribution of $X$.

For distributions with simple, explicit probability mass functions, the $p$-thinning can often be found directly, e.g., the $p$-thinning of a Geometric distribution with mean $\mu$ is Geometric with mean $p\mu$. 
However, when there is no closed-form expression for the probability mass function of $X$, the distribution of $Y$ can often be usefully approximated using saddlepoint methods. The CGF of $Y$ is guaranteed to exist if the CGF of $X$ exists: $K_Y(t) = K_X \{p \exp(t) + 1-p\}.$ 
We illustrate an implementation of a thinning-based model in \secref{sec:RSS_BernoulliRV}.

A related concept is multinomial splitting, where items counted by $X$ are assigned to several categories according to multinomial probabilities. Again, the multivariate CGF for the resulting counts is guaranteed to be available
as long as the CGF of $X$ is available.

\subsection{Randomly stopped sum}
\label{sec:RSS_section}

Consider variables of the form
\begin{equation}\label{eq:RSS}
    \boldsymbol{Y} = \sum_{i = 1}^{N} \boldsymbol{X}_i,
\end{equation}
where the $\boldsymbol{X}_i$ are i.i.d.\ and independent of the non-negative integer-valued random variable $N$. This construction is known as a randomly stopped sum.
Computing $\mathbb{P}(\boldsymbol{Y} = \boldsymbol{y})$ involves summing over all values of $N$ and the summands that are compatible with $\boldsymbol{y}$, and is usually intractable. 
However, if the CGFs of $N$ and the $\boldsymbol{X}_{i}$ are known, we can obtain the CGF of $\boldsymbol{Y}$ as $K_Y(\boldsymbol{t}) = K_N \{ K_X(\boldsymbol{t}) \}$,
enabling application of the saddlepoint likelihood method.

An example of this setup is the work of \cite{Meraou2022} to model aggregate claims from an insurance portfolio. 
The authors proposed models for the total number of claims, $N$, and individual claim amounts, $T_i$.
Noting that some claims may be denied, and only the sum of the approved payments is recorded, the observed aggregate claim $Y$ is
\begin{equation*}
    Y = \sum_{i = 1}^{L} T_i,~\text{ with}~L = \sum_{j=1}^{N} X_j,
\end{equation*}
where $X_j \sim \mathrm{Bernoulli}(p)$
specifies whether a claim was approved. Observations are available only when $L \geq 1$.
Here, the aggregate claim $Y$ is a randomly stopped sum, and $L$ is an instance of a thinning operation on the unobservable $N$ (see \secref{sec:Thinning}).
\cite{Meraou2022} estimated the underlying parameters using an expectation-maximization algorithm, but
the saddlepoint likelihood method provides an alternative approach because the CGF of $Y$ is available even though the distribution of $Y$ has no explicit form.
The CGF can be obtained iteratively 
by first addressing
the thinning process ($N$ to $L$) and then the randomly stopped sum ($L$ and $T_i$), such that
the output of the first CGF building block is the input for the second.
We provide a complete implementation of this example in \secref{CustomCGF} of the supplementary materials. 
For a simpler demonstration of a randomly stopped sum, see Sections \ref{sec:RSS_BernoulliRV} and \ref{sec:ReusingModelComponents}.

\subsection{Compound distributions}
Compound distributions are formed by combining two or more probability distributions, where one distribution typically generates parameter values for the other. 
For example, consider a scenario where counts follow a Poisson distribution, but the rate parameter is itself a random variable, $X$. We denote this as $(Y\,|\, X) \sim \mathrm{Poisson}(X)$.  
A Gamma distribution for $X$ leads to the Poisson-Gamma (or Negative Binomial) distribution with a known probability mass function, but other choices for $X$ can render exact calculations intractable. 
However, because the Poisson family of distributions is closed under addition, for integer-valued $X$
we can express $Y$ as a randomly stopped sum: $Y = \sum_{i = 1}^X P_i$, where $P_i \sim \mathrm{Poisson}(1)$.
This formulation aligns with the setup in \secref{sec:RSS_section}, for which we can obtain the CGF of $Y$ as long as the distribution of $X$ is known.
Crucially, the CGF operation that encodes the sum in \secref{sec:RSS_section} remains valid for $\mathrm{Poisson}(X)$ even for non-integer $X$. This flexibility allows us to obtain the CGF for $Y$ as $K_Y(t) = K_X \{ K_P(t) \}$ and apply saddlepoint methods regardless of whether $X$ is integer-valued or not.

The ability to obtain a CGF extends beyond the Poisson distribution.
If $X$ follows an arbitrary known distribution, and $(Y \, \vert \, X) \sim \mathrm{Binomial}(X,p)$ or $(Y \, \vert \, X) \sim \mathrm{Gamma}(X,\lambda)$, we can leverage the randomly stopped sum formulation to derive the corresponding CGFs.    
Notably, these distributions belong to families that are closed under addition, offering a clue for identifying other potential scenarios where we may successfully obtain CGFs.

\subsection{Sum of inhomogeneous random variables}\label{sec:sum_of_inhomogeneous_rvs}
Summing independent but non-identical random variables is another operation for which we can obtain the CGF. Computation of the exact probability distribution for these sums can be challenging. For instance, \cite{Liu2018} noted that the the exact distribution of a sum of independent $\mathrm{Binomial}(n_i, q_i)$ distributions is computationally prohibitive if there are more than two distinct $q_i$ values.

In a setting of parameter estimation, if the only available observation is the overall sum, the saddlepoint likelihood is a convenient option because the CGF of a sum of independent random variables is the sum of the component CGFs. In this context, \cite{Pedeli2015} considered an integer autoregressive process of order $p$. 
They expressed the integer-valued analogue of an AR($p$) process for a time variable $\tau \in \mathbbm{N}$ as
\[
Y_{\tau} =  \mathrm{Binomial}(Y_{\tau-1}, q_1) + \dotsb + \mathrm{Binomial}(Y_{\tau-p}, q_p)  + \epsilon_\tau.
\]
Here, the terms $\mathrm{Binomial}(Y_{\tau-i}, q_i)$ model the influence of $Y_{\tau-i}$ on the value of $Y_\tau$, and the $q_i$ are  analogues of the usual autoregressive coefficients. 
The innovations, $\epsilon_\tau$, are drawn from an integer-valued distribution.
\cite{Pedeli2015} demonstrated that the saddlepoint likelihood method provides a convenient and accurate estimation approach in a range of realistic settings.

\subsection{Partial summary}\label{sec:partial_summary}

Consider an observable random vector $\boldsymbol{Y}$ known to be a partial summary of some latent variables $\boldsymbol{X}$. The goal is to make inferences on the underlying parameters of the 
latent variables $\boldsymbol{X}$, without allowing the realized values of $\boldsymbol{X}$ to be discoverable.
In such cases, the observable counts $\boldsymbol{Y}$ typically have a complicated correlation structure, making exact calculations of probability mass functions intractable.
However, by leveraging the deterministic relationship between $\boldsymbol{X}$ and $\boldsymbol{Y}$, we may be able to use the saddlepoint likelihood for $\boldsymbol{Y}$ to uncover the underlying parameters of $\boldsymbol{X}$.

Partial summaries are common in demographic and health surveys, where
datasets may be released in an incomplete form due to privacy concerns.
For example, consider a health survey with variables including gender, smoking status, and diabetic status. A dataset of partial summaries might include the number of male smokers, number of male diabetics, and number of diabetic smokers, but omit the three-way count of male diabetic smokers which has a greater risk of compromising anonymity.
\cite{Dobra2006} explored this idea for inferences on multi-way contingency tables for which the data are supplied as marginal totals  across multiple subsets of variables. Following their work, \cite{Zhang2019} presented an illustration using the saddlepoint likelihood.  
Additional details on how to implement partial summaries are provided in \secref{sec:lin-transform-YAX}.

\subsection{Correlated count variables}\label{sec:Correlated_counts}

Generalizing from the setting in the previous section, any scenario involving count variables with shared latent components is likely to result in observed variables with elaborate multivariate dependence structures for which no explicit parametric form is available.
For instance, 
consider independent Poisson random variables $X_j$ with rates $\lambda_j$ for $j=0,\ldots,d$.
Define the following variables:
\begin{equation*}
    \begin{aligned}
         & Y_1 = X_1 + X_0\\
         & Y_2 = X_2 + X_0\\
         & ~~\vdots\\
         & Y_d = X_d + X_0.
    \end{aligned}
\end{equation*}
Since $X_0$ is common to all terms, the observable random vector $\boldsymbol{Y} = (Y_1, \ldots, Y_d)$ exhibits dependence among its components. 

Although the exact joint probability mass function of $\boldsymbol{Y}$ can be obtained in this simple case, calculations can quickly become computationally demanding in other models  \citep{Dimitris2003}.
However, if the relationship between $\boldsymbol{Y}$ and $\boldsymbol{X} = (X_0, \ldots, X_d)$ is linear, as in the example above, the CGF is readily available.
Specifically, if each component $Y_i$ can be written as $Y_i = \sum_j a_{ij} X_j$ for constants $(a_{ij})_{i=1,...,d; j=0,...,r}$, then $K_Y(t_1,...,t_d) = K_X(\tau_0, ..., \tau_r)$ where $\tau_j = \sum_i t_i a_{ij}$ for each $j$.
We provide an implementation of this example in \secref{sec:MVPoisson}.

\subsection{Capture-recapture with latent identities}\label{sec:latent_identities}

Capture-recapture describes a class of models for estimating population size and demographic parameters by repeatedly identifying individuals detected from the population. The population of interest might consist of humans, animals, or other objects. A common problem in capture-recapture studies is latent identities, where individuals cannot be uniquely distinguished from others. For example, \cite{Zhang2022} used batch-mark data, in which recaptured animals are known to belong to a batch-marked group, but the particular animals are not distinctively known. Latent identities can also occur when an individual's identity manifests in different ways. This is seen in certain models for misidentification \citep{Link2010} and in studies that employ more than one method for identifying animals, such as photographs and DNA samples \citep{Bonner_Holmberg2013}.

Capture-recapture modeling typically proceeds by constructing counts of the numbers of individuals with each distinct encounter history. In the latent identity context, these count vectors $\boldsymbol{X}$ cannot be observed, but rather they are mapped onto an observable vector $\boldsymbol{Y}$, for which the probability mass function is intractable. 
However, the mapping from $\boldsymbol{X}$ to $\boldsymbol{Y}$ is typically linear, providing access to the CGF of $\boldsymbol{Y}$ as in the previous section, and enabling application of the saddlepoint likelihood method.
Details are given in \secref{sec:lin-transform-YAX}, and an implementation of a latent identity model is provided in \secref{sec:Two_source_example}.

\subsection{\texorpdfstring{Linear transformation $\boldsymbol{Y=AX}$}{Linear transformation Y=AX}
}
\label{sec:lin-transform-YAX}

The examples discussed in Sections \ref{sec:partial_summary}, \ref{sec:Correlated_counts}, and \ref{sec:latent_identities} are instances of a general type of problem known as the statistical linear inverse model \citep{Hazelton_Y_AX_2020}.
In this general formulation, we relate our observable vector of counts, $\boldsymbol{Y} = (Y_1,\ldots,Y_d)^T$, to an unobservable count vector of interest, $\boldsymbol{X} = (X_1,\dots,X_r)^T$, through the linear transformation $\boldsymbol{Y} = 
\boldsymbol{AX}$. 
Here, $\boldsymbol{A}$ is a $(d \times r)$ deterministic matrix, typically non-invertible.
The structure of the matrix $\boldsymbol{A}$ encodes specific relationships based on the context.
For the partial summary in Section \ref{sec:partial_summary}, the rows of $\boldsymbol{A}$ (composed of zeros and ones) indicate which attributes contribute to the partial totals. In the capture-recapture problems in Section \ref{sec:latent_identities}, the columns of matrix $\boldsymbol{A}$ encode the correspondence of the elements of the observed vector $\boldsymbol{Y}$ to different latent identities. 

Linear inverse models arise in a large and diverse range of applications. As demonstrated by \cite{Zhang2019}, the saddlepoint likelihood method offers an effective estimation framework for this model class. For argument $\boldsymbol{t} \in \mathbb{R}^d$, the CGF of $\boldsymbol{Y}$ is $K_Y(\boldsymbol{t}) = K_X(\boldsymbol{tA})$, where as usual $\boldsymbol{t}$ is interpreted as a row vector.
See \secref{sec:Two_source_example} for an example.

\subsection{Validation of the saddlepoint likelihood method}\label{sec:validation}

Sections~\ref{sec:sum-of-iid}--\ref{sec:lin-transform-YAX} illustrate how the saddlepoint approximation enables likelihood-based inference for a broad class of models with otherwise intractable likelihoods. In each case, however, the likelihood being maximized is an approximation, albeit typically a highly accurate one.
To assist with validating the saddlepoint approach, we define the \textit{discrepancy} to be the difference between the MLEs obtained from the saddlepoint and exact likelihoods.
As part of our framework, we provide an explicit quantification of the discrepancy            
via an accurate approximation that is computable irrespective of whether the exact MLE is available.
We provide computational details in \secref{sec:discrepancy} and theoretical development in \citet{oketch2025}.  
The discrepancy approximation can be evaluated without access to the exact likelihood because it relies only on the CGF building blocks described in Sections~\ref{sec:sum-of-iid}--\ref{sec:lin-transform-YAX}; no additional model-specific derivations are required.

For practical validation of the saddlepoint likelihood method we can compare the size of the discrepancy to the magnitude of sampling variability.
In examples in \secref{sec:Examples}, we show that the discrepancy is typically a small fraction of the standard error, indicating that the error introduced by the saddlepoint approximation is negligible.

\section{Computational framework}\label{sec:FrameworkSection}

Based on \eqref{eq:Saddlepoint_Defn} and \eqref{eq:Likelihood_Fn}, we define the saddlepoint log-likelihood function as follows:
\begin{equation}\label{eq:Log-likelihood-Multivariate}
    \begin{aligned}
          & \log \hat{L} (\boldsymbol{\theta}\, | \, \boldsymbol{y}) = K_{Y}(\hat{\boldsymbol{t}};\boldsymbol{\theta}) - \hat{\boldsymbol{t}}\boldsymbol{y} - \frac{d}{2} \log(2\pi) - \frac{1}{2} \log \det ( K_{Y}''(\hat{\boldsymbol{t}};\boldsymbol{\theta}) ),
    \end{aligned}
\end{equation}
where $\hat{\boldsymbol{t}} = \hat{\boldsymbol{t}}(\boldsymbol{\theta};\boldsymbol{y})$ is the solution of $K_{Y}'(\hat{\boldsymbol{t}};\boldsymbol{\theta})=\boldsymbol{y}$, as in \eqref{eqn:saddlepoint_eqn}.

We can compute \eqref{eq:Log-likelihood-Multivariate} when we have access to $K_{Y}$, the CGF of $\boldsymbol{Y}$. 
As highlighted in Section~\ref{sec:ModelBuildingBlocks}, obtaining $K_{Y}$ typically involves operations on the CGFs of other variables with known distributions. 
These operations, although individually straightforward, can become complicated when combined.
In Section \ref{sec:sum_of_inhomogeneous_rvs}, for instance, thinning and inhomogeneous sum operations are combined to define the autoregressive process, and several operations are also required for the example in \secref{sec:RSS_section}.
It quickly becomes cumbersome to determine the functions $K_Y$, $K_Y'$, and $K_Y''$ required in \eqref{eq:Log-likelihood-Multivariate} when dealing with multiple successive operations. 

In our framework, we automatically compute the requisite CGFs
and their derivatives by systematically translating modeling problems involving the operations from \secref{sec:ModelBuildingBlocks} into their corresponding CGFs. 
This eliminates the need for users to engage with the mathematical details of $K_Y$, $K_Y'$, $K_Y''$, and the evaluation and optimization of \eqref{eq:Log-likelihood-Multivariate}.
The framework also accommodates user-defined CGFs; an illustrative example is provided in \secref{CustomCGF} of the supplementary materials.

The framework is designed for use with R \citep{R2022}. 
We represent a CGF as a code object that implements the function $K(\boldsymbol{t};\boldsymbol{\theta})$, related functions such as $K'(\boldsymbol{t};\boldsymbol{\theta})$ and $K''(\boldsymbol{t};\boldsymbol{\theta})$, and other special-purpose operations.
When we apply a mathematical operation to transform a known distribution $\boldsymbol{X}$ into a new distribution $\boldsymbol{Y}$, we represent this by invoking a function that accepts the CGF code object for $\boldsymbol{X}$ and returns a new code object for $\boldsymbol{Y}$. This process can be iterated using the code object for $\boldsymbol{Y}$ as input to subsequent operations.

\subsection{CGF code objects for generic model-building}
\label{sec:cgf-object}

A mathematical representation of a CGF is a single function $K_Y:\mathbb{R}^{n} \to \mathbb{R}$ for a fixed random vector $\boldsymbol{Y}$ of fixed dimension $n$. 
Our CGF code objects are intentionally more general.
Each object represents a family of parametric distributions via a map $(\boldsymbol{t}, \boldsymbol{\theta}) \mapsto K_Y(\boldsymbol{t}; \boldsymbol{\theta})$. 
In addition, objects may accept $\boldsymbol{t}$-vectors of varying lengths, with longer inputs treated automatically as i.i.d.\ replicates.
Thus a single object can represent both a multivariate distribution $\boldsymbol{Y}$ depending on a parameter vector $\boldsymbol{\theta}$, and any number of i.i.d.\ replicates of that distribution.

The dimension $d$ of $\boldsymbol{Y}$ and the number $m$ of i.i.d.\ replicates are determined at evaluation time, in a distribution-appropriate manner, from the lengths of the supplied vectors $\boldsymbol{t}$ and $\boldsymbol{\theta}$. Alternatively, the user may fix $d$ and $m$ when constructing the object.  
When $d$ and $m$ are fixed, the object behaves like a classical CGF $K_Y:\mathbb{R}^{n} \to \mathbb{R}$ for a parametric distribution of dimension $n=md$.
We explain this further in \secref{sec:vec-dimension}, with specific examples in Sections~\ref{sec:MVPoisson}--\ref{sec:RSS_BernoulliRV}.

Similar to creating an R function, creating a CGF object does not immediately trigger its evaluation.
In particular, dimension and validity checks are only enforced at evaluation time once $\boldsymbol{t}$ and $\boldsymbol{\theta}$ are supplied.  
However, some later top-level operations will cause CGF objects to be evaluated automatically, typically in the context of an optimizer or an automatic differentiation taping operation, both of which require a fixed dimension.

Every CGF code object, whether it represents a base distribution such as Poisson or multinomial, or a distribution obtained by chaining model‑building operations as in \secref{sec:ModelBuildingBlocks}, exposes the same public interface. This interface comprises $K_Y(\boldsymbol{t};\boldsymbol{\theta})$, $K'_Y(\boldsymbol{t};\boldsymbol{\theta})$, $K''_Y(\boldsymbol{t};\boldsymbol{\theta})$, a solver for the saddlepoint equation $K'_Y(\boldsymbol{t};\boldsymbol{\theta}) = \boldsymbol{y}$, and the operator functions used for our discrepancy diagnostic (\secref{sec:discrepancy}).

\subsection{CGF operations}
\label{sec:CGFoperations}
The modeling operations in \secref{sec:ModelBuildingBlocks} are transformations of random variables, which induce transformations of the corresponding CGFs through the argument $\boldsymbol{t}$.
For each of these operations we provide a construction that takes a CGF object for $\boldsymbol{X}$ and returns a CGF object for the transformed variable $\boldsymbol{Y}$.
The construction manages the full $\boldsymbol{t}$‑dependence of $K_Y$, including its first and second derivatives and any restrictions on the domain of $\boldsymbol{t}$.

Because every construction returns a CGF object with the same public interface, operations can be chained. For instance, a linear mapping can be applied to a sum of independent terms, which can in turn be embedded inside a randomly stopped sum, and so on.
\figref{fig:Flowchart1} provides a schematic of this assembly process.

\subsection{Adaptor functions}
\label{sec:adaptor_function}

A full specification of a modeled variable $\boldsymbol{Y}$ requires a mapping from the model parameters to the distributional parameters of the variables involved in constructing $\boldsymbol{Y}$. 
Suppose the variable $\boldsymbol{X}$ has a known distribution 
with parameter $\boldsymbol{\varphi}$, which we term the {\em  distributional parameter}. 
For instance, if $\boldsymbol{X}$ is multinomial the index $N$ and probability vector $\vec{\boldsymbol{p}}$ yield  $\boldsymbol{\varphi} = (N,\vec{\boldsymbol{p}})$.
The distributional parameters are typically determined by a smaller set of model parameters, which we denote by $\boldsymbol{\theta}$.
We write $\boldsymbol{X} = \boldsymbol{X}_{\boldsymbol{\varphi}}$ and $\boldsymbol{Y} = \boldsymbol{Y}_{\boldsymbol{\theta}}$, and define
\begin{equation}\label{eq:Link_adaptor_fn}
     \boldsymbol{\varphi} = h(\boldsymbol{\theta}).
\end{equation}
The function $h$ in \eqref{eq:Link_adaptor_fn} is called an {\em adaptor} linking the model parameters to the distributional parameters. (Our use of the term ``adaptor'' borrows from its everyday meaning as a device that connects a specific appliance to a standard mains power supply.) 
Because CGF objects represent parametric distributions, 
each object must specify the structure of the parameter vector it expects.

Our approach is to frame the modeling process in terms of two coding components: CGF operations and adaptors.
This allows us to preprogram CGF objects for a spectrum of common distributions $\boldsymbol{X}$ with respect to their distributional parameters, then allow these CGFs to be tailored via adaptors to yield the CGF for any model of interest. 
Adaptors can be handled generically, along with any parametric constraints, using automatic differentiation to compute gradients \citep{TMB2016}. 
For elaborate models, we might require a suite of adaptors to link the model parameters to different distributional parameters.

Adaptor functions typically arise naturally from the specification of the model: see for instance the capture-recapture model in \secref{sec:Two_source_example}, where the multinomial parameters $N$ and $\vec{\boldsymbol{p}}$ are expressed in terms of population size and per-occasion capture probabilities.
In our software implementation, they achieve the transformation from a generic CGF object of $\boldsymbol{X}$, which includes \{$K_X(\boldsymbol{t};\boldsymbol{\varphi})$, $K'_X(\boldsymbol{t};\boldsymbol{\varphi})$, $K''_X(\boldsymbol{t};\boldsymbol{\varphi}) $\},
to the model-specific CGF object of $\boldsymbol{Y}$, containing \{$K_Y(\boldsymbol{t};\boldsymbol{\theta})$, $K'_Y(\boldsymbol{t};\boldsymbol{\theta})$, $K''_Y(\boldsymbol{t};\boldsymbol{\theta}) $\}. 
By expressing the CGF object in terms of differentiable functions of the model parameter $\boldsymbol{\theta}$, we can automate the gradient-based optimization of \eqref{eq:Log-likelihood-Multivariate} to obtain estimates of $\boldsymbol{\theta}$.

\subsection{Modeling workflow}\label{sec:workflow}

\begin{figure}[hbtp]
\centering
\includegraphics[width=0.95\linewidth]{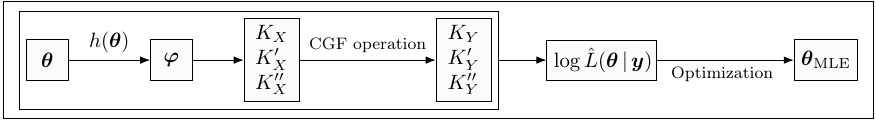} 
\caption{
Schematic showing how we assemble the CGF of $\boldsymbol{Y}$, using an adaptor function $h(\boldsymbol{\theta})$ as in \secref{sec:adaptor_function} and a CGF operation as \secref{sec:CGFoperations}.
Once the desired CGF object is built, the saddlepoint log-likelihood and its gradient are calculated automatically. A single function invocation can then be used to find the saddlepoint MLE.
}
\label{fig:Flowchart1}
\end{figure}

\figref{fig:Flowchart1} summarizes a typical workflow. Starting from a base CGF object for
$\boldsymbol{X}$, one or more CGF operations map it to the object for $\boldsymbol{Y}$ (\secref{sec:CGFoperations}),  while adaptors are used to link model parameters to distributional parameters when required (\secref{sec:adaptor_function}).

We implement our computational framework in the R package \texttt{saddlepoint}, available from \texttt{remotes::install\_github("godrick/saddlepoint")}. 
The package provides implementations and examples for each of the CGF operations described in \secref{sec:ModelBuildingBlocks}.
For example, if the CGF for $\boldsymbol{X}$ is stored as R object \texttt{K.X}, then \texttt{K.Y <- linearlyMappedCGF(K.X, A)}
constructs the CGF code object for the random vector $\boldsymbol{Y=AX}$. For summing $n$ i.i.d\, copies of $\boldsymbol{X}$ such that $\boldsymbol{Y}=\boldsymbol{X}_1+\dots+\boldsymbol{X}_n$, we use \texttt{K.Y <- sumOfiidCGF(K.X, n)}.
In addition, CGF objects may incorporate one
or more adaptors.
Creating the CGF code object for a parametric model based on a core distribution and an adaptor is made easy by functions such as \texttt{BinomialModelCGF} and \texttt{PoissonModelCGF}, which take the adaptor function as an argument and return the corresponding CGF code object.
See Sections \ref{sec:MVPoisson}--\ref{sec:ReusingModelComponents} for illustrations.

When CGF objects are constructed using our preprogrammed suite and built-in operations, all subsequent computations are handled internally.
In particular, optimization of the saddlepoint likelihood, including assembly of the automatic differentiation objects, can be achieved by a single call to the function \texttt{find.saddlepoint.MLE()}.
Additional top level functions include \texttt{compute.spa.negll()}, which evaluates the saddlepoint log‑likelihood for a CGF object without performing any optimization.
This can be used, for example, to combine saddlepoint likelihood terms with other likelihood terms prior to optimization.

\subsection{Discrepancy computation}\label{sec:discrepancy}

Maximizing the saddlepoint log-likelihood $\log \hat{L}(\boldsymbol{\theta}\;|\;\boldsymbol{y})$ instead of an exact log-likelihood $\log L(\boldsymbol{\theta}\;|\;\boldsymbol{y})$ induces a discrepancy between the resulting MLEs.
We define this discrepancy as 
$
    \boldsymbol{\delta} = \hat{\boldsymbol{\theta}}_{\mathrm{exact}} - \hat{\boldsymbol{\theta}}_{\mathrm{spa}},
$
where
$
    \hat{\boldsymbol{\theta}}_{\mathrm{exact}} 
    =
    \mathrm{argmax}_{\theta}\, \log L(\boldsymbol{\theta}\;|\;\boldsymbol{y})
$
and
$
    \hat{\boldsymbol{\theta}}_{\mathrm{spa}} 
    =
    \mathrm{argmax}_{\theta}\, \log \hat{L}(\boldsymbol{\theta}\;|\;\boldsymbol{y}).
$
\citet[Definition~2]{oketch2025} provide an accurate closed‑form approximation to $\boldsymbol{\delta}$:
\[
  \hat{\boldsymbol{\delta}}
    = -\bigl\{\nabla_{\theta}^2\log
      \hat L(\hat{\boldsymbol\theta}_{\mathrm{spa}}\mid\boldsymbol y)\bigr\}^{-1}
      \nabla_{\theta} T_Y(\hat{\boldsymbol\theta}_{\mathrm{spa}},\boldsymbol y),
\]
where $T_Y(\boldsymbol{\theta},\boldsymbol{y})$ is the correction term to the first‑order saddlepoint log‑likelihood,
\begin{equation}\label{eq:Tcorr}
\begin{aligned}
T_Y(\boldsymbol{\theta},\boldsymbol{y})
&= \frac18
     \sum_{j_1,j_2,j_3,j_4}^{d}
       K^{(4)}_{j_1j_2j_3j_4}\,
       Q_{j_1j_2}Q_{j_3j_4} \\
&\quad 
     -\frac18
     \sum_{j_1,\dots,j_6}^{d}
       K^{(3)}_{j_1j_2j_3}\,
       K^{(3)}_{j_4j_5j_6}\,
       Q_{j_1j_2}Q_{j_3j_4}Q_{j_5j_6} \\
&\quad
     -\frac1{12}
     \sum_{j_1,\dots,j_6}^{d}
       K^{(3)}_{j_1j_2j_3}\,
       K^{(3)}_{j_4j_5j_6}\,
       Q_{j_1j_4}Q_{j_2j_5}Q_{j_3j_6},
\end{aligned}
\end{equation}
with $Q = K_Y''(\hat {\boldsymbol{t}}(\boldsymbol{\theta};\boldsymbol{y});\boldsymbol{\theta})^{-1}$ and where
$K^{(r)}$ denotes the $r$‑th derivative tensor of the CGF evaluated at
the saddlepoint $\hat {\boldsymbol{t}}(\boldsymbol{\theta};\boldsymbol{y})$.

In our framework, the gradient
$
 \nabla_{\theta}T_Y(\hat{\boldsymbol\theta}_{\mathrm{spa}},\boldsymbol y)
$
is obtained using automatic differentiation via the R package RTMB \citep{TMB2016}. 
However, evaluating 
$
 T_Y(\boldsymbol\theta,\boldsymbol y)
$
naively, by computing arrays of third‑ or fourth‑order partial derivatives,
can be computationally slow. For details, see \citet[Section~4]{oketch2025}.
Instead we adopt the approach in \citet[Supplement~5]{oketch2025}. 
This involves several specialized functions that perform tensor contractions on the fly, so no third‑ or fourth‑order derivative arrays are ever materialized in memory.
The diagnostic is obtained simply by enabling the option \texttt{discrepancy = TRUE} in \texttt{find.saddlepoint.MLE()}. The CGF code object described in \secref{sec:cgf-object} is used unchanged. Examples illustrating this option appear in \secref{sec:Examples}.

\subsection{Handling vector dimensions and i.i.d.\ replication
}\label{sec:vec-dimension}

For a random vector $\boldsymbol{Z}=(Z_1, \dots, Z_m)$ consisting of $m$ i.i.d.\ replicates, the CGF argument \(\boldsymbol t=(\boldsymbol t^{(1)},\dots,\boldsymbol t^{(m)})\) partitions naturally into $m$ blocks, such that
\begin{equation}\label{eq:m-iid-CGF-Z}
    K_Z(\boldsymbol{t}) = K_{Z_1}(\boldsymbol t^{(1)}) + \dots + K_{Z_m} (\boldsymbol t^{(m)}).
\end{equation}
If each block $\boldsymbol{Z}_i$ itself has dimension $d$, then each block $\boldsymbol t^{(i)}$ will also have dimension $d$. 
More generally, random vectors may comprise multiple building blocks with different dimensions.

In our framework, the argument $\boldsymbol t$ is always passed as a single vector.
Consequently, CGF objects must be able to interpret the argument $\boldsymbol t$ and partition it into appropriate subblocks.
Most CGF objects in our framework behave as follows: if a vector $\boldsymbol t$ of length $n$ is supplied, it is interpreted as $n = md$, where $d$ is the block size implied by the distribution and $m$ is arbitrary.
The corresponding random vector $\boldsymbol Z$ is then interpreted as $m$ i.i.d.\ replicates of block size $d$.
For example, CGF objects for scalar distributions always select $d = 1$.
An exception is the CGF object for the multinomial distribution, which selects $d$ dynamically based on the length of the multinomial probability vector supplied in the accompanying parameters.  

Users can override the default behavior
via the optional arguments {\tt block\_size = d} and {\tt iidReps = m} when setting up a CGF object. In this case, subsequent calls to that object will raise an error if incompatible lengths are supplied. This can be useful for error checking.

Certain CGF operations, such as {\tt concatenationCGF()} and {\tt randomlyStoppedSumCGF()}, require additional disambiguation of block size or number of replicates. This is explained in the documentation for the corresponding operations.

We remark that CGF objects do not exhibit vectorization as commonly seen in R functions, even when a vector argument $\boldsymbol t$ is interpreted as $m$ subblocks corresponding to i.i.d. replicates. 
For instance, the CGF function $K_Z(\boldsymbol{t})$ from \eqref{eq:m-iid-CGF-Z} always returns a scalar, not a vector of CGF values for each $\boldsymbol{t}^{(i)}$.

\section{Constrained optimization approach}
\label{sec:ConstrainedOptim}

We adopt a constrained optimization strategy for maximizing the log-likelihood function \eqref{eq:Log-likelihood-Multivariate}.
This approach differs from the two-step optimization approach that has traditionally been used for saddlepoint estimation, in which an outer optimization takes place over $\boldsymbol{\theta}$, and for each $\boldsymbol{\theta}$ an inner optimization step is used to find the saddlepoint $\hat{\boldsymbol{t}}(\boldsymbol{\theta}; \boldsymbol{y})$ \citep{Zhang2019, TMB2016}. By contrast, the constrained approach optimizes simultaneously over $\boldsymbol{\theta}$ and $\boldsymbol{t}$, 
which are constrained to satisfy the saddlepoint equation. This has certain advantages as we outline below.

Consider the following general formulation of a constrained optimization problem involving equality and inequality constraints: 
\begin{equation*}\label{eq:General_Defn_NLOPT}
    \begin{aligned}
         & {\mathrm{maximize}}~g(\boldsymbol{z}),\\
         & \text{subject to:}~c_\mathrm{eq}(\boldsymbol{z}) = 0 \\
         & \quad\quad\quad\quad\quad c_\mathrm{ineq}(\boldsymbol{z}) \leq 0,
    \end{aligned}
\end{equation*}
where $\boldsymbol{z} \in \mathbb{R}^{n}$, $g: \mathbb{R}^{n} \to \mathbb{R}$, $c_\mathrm{eq}: \mathbb{R}^{n} \to \mathbb{R}^d$ and $c_\mathrm{ineq}: \mathbb{R}^{n} \to \mathbb{R}^r$. For the saddlepoint log-likelihood function, $\boldsymbol{z}$ defines a combined row vector of the saddlepoints $\boldsymbol{t} \in \mathbb{R}^d$ and the model parameter vector $\boldsymbol{\theta} \in \mathbb{R}^p$, i.e., $\boldsymbol{z} = (\boldsymbol{t}, \boldsymbol{\theta}) \in \mathbb{R}^{n}$. 
Therefore, our formulation is
\begin{equation}\label{eq:Saddlepoint_Defn_NLOPT}
    \begin{aligned}
         & {\mathrm{maximize}}~g(\boldsymbol{t},\boldsymbol{\theta})\\
         & \text{subject to:}~c_\mathrm{eq}(\boldsymbol{t},\boldsymbol{\theta}) = 0\\
         & \quad\quad\quad\quad\quad c_\mathrm{ineq}(\boldsymbol{t},\boldsymbol{\theta}) \leq 0,
    \end{aligned}
\end{equation}
where the objective function and equality constraints are specified by \eqref{eq:Log-likelihood-Multivariate} and \eqref{eqn:saddlepoint_eqn},
\begin{equation*}
    \begin{aligned}
         g(\boldsymbol{t},\boldsymbol{\theta}) 
         &= K_Y(\boldsymbol{t};\boldsymbol{\theta}) - \boldsymbol{t} K_Y'(\boldsymbol{t};\boldsymbol{\theta}) - \frac{d}{2} \log(2\pi) - \frac{1}{2} \log \det \{ K_Y''(\boldsymbol{t};\boldsymbol{\theta}) \},\\
         c_\mathrm{eq}(\boldsymbol{t},\boldsymbol{\theta}) 
         &= K_Y'\left(\boldsymbol{t};\boldsymbol{\theta}\right)-\boldsymbol{y}, 
    \end{aligned}
\end{equation*}
and $c_\mathrm{ineq}(\boldsymbol{t},\boldsymbol{\theta})$ is an optional function that incorporates any model-specific constraints on the parameter space involving $\boldsymbol{\theta}$, constraints in the domain of $\boldsymbol{t}$, or both. 

There are several constrained optimization algorithms that accommodate equality and inequality constraints: see \cite{Boyd2004} and \cite{Nlopt} for a summary. 
Using any of these algorithms on \eqref{eq:Saddlepoint_Defn_NLOPT}, 
we simultaneously update $\boldsymbol{t}$ and $\boldsymbol{\theta}$ at each iteration. 
Because the objective function $g(\boldsymbol{t}, \boldsymbol{\theta})$ and the constraints are explicit functions of $\boldsymbol{t}$ and $\boldsymbol{\theta}$, their gradients are straightforward to compute. We write  
\begin{equation}
\label{eq:grad_z}
    \spacednabla{\boldsymbol{z}}{g} = (\spacednabla{\boldsymbol{t}}{g}, \spacednabla{\boldsymbol{\theta}}{g}),
    \quad\quad
    \spacednabla{\boldsymbol{z}}{c_\mathrm{eq}(\boldsymbol{t},\boldsymbol{\theta})} = \Big(K_Y''(\boldsymbol{t};\boldsymbol{\theta}), \nabla_{\boldsymbol{\theta}} K_Y'(\boldsymbol{t};\boldsymbol{\theta})\Big),
\end{equation}
such that for $\boldsymbol{t} \in \mathbb{R}^d$ and $\boldsymbol{\theta} \in \mathbb{R}^p$, $\spacednabla{\boldsymbol{z}}{g}$ concatenates row vectors $\spacednabla{\boldsymbol{t}}{g}$ and $\spacednabla{\boldsymbol{\theta}}{g}$ of size $d$ and $p$, respectively. The gradient $\nabla_{\boldsymbol{z}} c_\mathrm{eq}(\boldsymbol{t},\boldsymbol{\theta})$ is a $d \times (d+p)$ block matrix.
Where an adaptor $h(\boldsymbol{\theta})$ has been used to create a CGF $K_Y$ from a parent CGF $K_X$, the gradient of $K_Y$ with respect to $\boldsymbol{\theta}$ is readily obtained from that of $K_X$ via the expression 
\begin{equation}
    \label{eq:grad_theta_K_with_adaptor}
    \nabla_{\boldsymbol{\theta}} K_Y(\boldsymbol{t};\boldsymbol{\theta}) = h'(\boldsymbol{\theta}) \nabla_{\boldsymbol{\theta}}K_X(\boldsymbol{t};h(\boldsymbol{\theta})).
\end{equation}

\begin{figure}[H]
\centering
\includegraphics[width=1.0\linewidth]{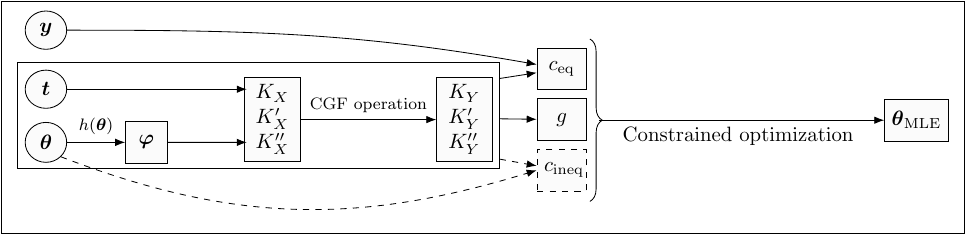} 
\caption{
Flowchart depicting the integration of constrained optimization into the framework shown in \figref{fig:Flowchart1}. The equality constraint $c_{\mathrm{eq}}$ depends on observations $\boldsymbol{y}$, with an optional inequality on $\boldsymbol{t}$ applied via the CGF. Additional model-imposed inequalities can further contribute to the function $c_{\mathrm{ineq}}$.
These components, along with the objective function, are processed by a constrained optimization algorithm to yield MLEs.
}
\label{fig:Flowchart2}
\end{figure}

By contrast, in the classical two-step approach to maximizing \eqref{eq:Log-likelihood-Multivariate}, the log-likelihood is a function of $\boldsymbol{\theta}$ both directly, and indirectly through $\hat{\boldsymbol{t}}$, so the computation of the $\boldsymbol{\theta}$-based gradients is 
more complicated. For details of this gradient computation and 
the two-step approach, see \secref{sec:two_step_approach} of the supplementary materials.
We favor the constrained approach for our general-purpose framework because it eliminates the implicitly defined functions and can easily incorporate additional inequality constraints on $\boldsymbol{t}$ and $\boldsymbol{\theta}$ as needed for user-defined models. 
\figref{fig:Flowchart2} illustrates how data and parameter inputs are combined with other model-specific constraints to produce saddlepoint MLEs in the constrained optimization framework.
The computation of \eqref{eq:grad_z} and \eqref{eq:grad_theta_K_with_adaptor} using automatic differentiation software \citep{TMB2016} facilitates robust and efficient convergence of the chosen optimization algorithm.

\vspace*{-10mm}
\section{Applications}
\label{sec:Examples}

In this section we illustrate 
the ease with which our
framework can be employed for practical modeling. Using examples we show how researchers can perform saddlepoint-based likelihood estimation solely by specifying model-facing components of a problem.
Our functions then internally assemble the saddlepoint likelihood and the necessary gradients to complete the estimation process.
All examples in \secref{sec:Examples} are implemented in full in code supplied in the supplementary materials, accompanied by additional explanatory guidelines for modelers.

\subsection{Multivariate Poisson model}\label{sec:MVPoisson}

Following from the example in \secref{sec:Correlated_counts}, suppose we observe $m$ i.i.d.\ copies of a vector $\boldsymbol{Y} = (Y_1,\ldots,Y_d)$ which follows a multivariate Poisson distribution such that $Y_i = X_i + Z_0$, where
$X_i$ and $Z_0$ are unobservable, independent Poisson variables with distributional parameters  $\alpha$ and $\beta$, respectively.
For specificity, we implement $m=5$ and $d=3$.

Expressed as vectors, each instance of $\boldsymbol{Y}$ can be written as 
\begin{equation}\label{eq:MVPois_SumIndependent_Example}
    \boldsymbol{Y} = \boldsymbol{X} + \boldsymbol{A} Z_0,
\end{equation}
such that $\boldsymbol{X} = (X_1,\ldots, X_d)$ is made up of i.i.d.\ Poisson variables, and $\boldsymbol{A}$ is a $d$-dimensional column vector of ones. The goal is to estimate the model parameter $\boldsymbol{\theta} = (\alpha, \beta)$ based on i.i.d.\ observations of the vector $\boldsymbol{Y}$.

In this simple example, the CGF of $\boldsymbol{Y}$ is available in explicit form, such that for $\boldsymbol{t} \in \mathbb{R}^d$,
\begin{equation}\label{eq:MVProblemIncludingTheta}
K_Y(\boldsymbol{t}; \boldsymbol{\theta}) = K_X(\boldsymbol{t}; \alpha) + K_{Z_0}(\boldsymbol{tA}; \beta).
\end{equation}
However, in our framework there is no need for manual derivation of \eqref{eq:MVProblemIncludingTheta} and its derivatives. 
Instead, the necessary CGF can be obtained exclusively by referencing the model structure in \eqref{eq:MVPois_SumIndependent_Example}.

To accomplish this, we first obtain
the CGFs corresponding to $\boldsymbol{X}$ and $\boldsymbol{A}Z_0$ using the \verb|PoissonModelCGF()| and \verb|linearlyMappedCGF()| functions (lines 4--9, \coderef{code:MVPoisson}). 
These functions are wrappers in R that invoke the relevant CGF object and the specific building block among those described in \secref{sec:ModelBuildingBlocks}.
To ensure that these CGFs are treated as functions of the model parameter $\boldsymbol{\theta}$, rather than the distributional parameter, the function \verb|adaptor(indices = ...)| is used. The \verb|indices| option identifies the positions of distributional parameters in the model parameter vector $\boldsymbol{\theta} = (\alpha, \beta)$. For $\boldsymbol{X}$, which depends on $\alpha$ (the first position of $\boldsymbol{\theta}$), we set \verb|indices = 1| in \coderef{code:MVPoisson}[4] of \coderef{code:MVPoisson}, while for $Z_0$ we set \verb|indices = 2| in \coderef{code:MVPoisson}[8]. The function \verb|PoissonModelCGF()| 
creates the CGF object in R, including the CGF, its derivatives, and adaptor information.

\begin{table}[hbtp]
\centering
\begin{lstlisting}[caption = CGF of the multivariate Poisson random variable in \secref{sec:MVPoisson}, label = code:MVPoisson, linewidth=1.03\textwidth]
# Specifying dimensions: 5 i.i.d. replicates of 3-dimensional vector
m = 5; d = 3
# CGF of X specifying Poisson with iidReps=d, so K.X expects an argument of length d 
K.X = PoissonModelCGF(lambda = adaptor(indices = 1), iidReps = d) 
# CGF of (A*Z0)
# A is dx1, so K.AZ0 expects an argument of length d
A = matrix(1, nrow = d, ncol = 1)
K.Z0 = PoissonModelCGF(lambda = adaptor(indices = 2), iidReps = 1) 
K.AZ0 = linearlyMappedCGF(cgf = K.Z0, matrix_A = A) 
# CGF corresponding to i.i.d. observations of vector Y
# An argument of length (md) will be partitioned into m blocks each of length d
K.Y_iid = sumOfIndependentCGF(list(K.X, K.AZ0), iidReps = m, block_size = d) 
\end{lstlisting} 
\end{table}

Next, we obtain the CGF of $\boldsymbol{Y} = \boldsymbol{X} + \boldsymbol{A}Z_0$ using the \texttt{sumOfIndependentCGF()} function based on the formulation in \eqref{eq:MVPois_SumIndependent_Example}. To extend the resulting CGF object to 
an object for $m$ i.i.d.\ copies of the $d$-dimensional vector $\boldsymbol{Y}$,
we specify \texttt{iidReps = m, block\_size = d} in \coderef{code:MVPoisson}[12]. 
Because we specify both \texttt{iidReps} and \texttt{block\_size}, the resulting CGF object \texttt{K.Y\_iid} expects a vector of length $md=15$ exactly.
This object can then be used in
\texttt{find.saddlepoint.MLE()} to obtain parameter estimates and, optionally, the discrepancy diagnostic defined in
Sections~\ref{sec:validation} and \ref{sec:discrepancy}:
\begin{verbatim}
    find.saddlepoint.MLE(observed.data = y, cgf = K.Y_iid,
                         starting.theta = c(1, 1),
                         discrepancy    = TRUE,
                         std.error      = TRUE)
\end{verbatim}
The observations $\texttt{y}$ are supplied as a vector of length $md$, where the first $d$ elements correspond to the first instance of the $d$-dimensional random variable $\boldsymbol{Y}$, and so on.

\coderef{code:MVPoisson} exemplifies how to build CGFs and prepare models for estimation using the saddlepoint likelihood,
using only model-facing operations. Contrasting with the customary approach, which 
involves manually deriving the CGF as in \eqref{eq:MVProblemIncludingTheta}, 
followed by the computation of its first and second derivatives,
the code snippets in \coderef{code:MVPoisson} provide an intuitive and streamlined way of constructing the CGF that aligns with the model's structural components. This approach reduces 
the whole process of saddlepoint likelihood estimation to a few concise lines of code.

For this simple multivariate Poisson problem, the exact probability mass function of $\boldsymbol{Y}$ is computationally tractable. 
Therefore, we compared the estimates based on the saddlepoint likelihood with those from the exact likelihood function.
These results are presented in \reffig{fig:MVPoisson}. 
The saddlepoint-based estimates closely match those from the exact likelihood, with discrepancies that are only a small fraction of the standard errors. 
The largest discrepancy among the 100 simulated datasets is about 20\% of the corresponding standard error. 
The plots of the standard errors in \reffig{fig:MVPoisson} also show close agreement, reflecting agreement in the corresponding Hessian matrices and further demonstrating the viability of using the saddlepoint likelihood as an alternative to the exact likelihood function.
The third row compares the true discrepancies with the approximated discrepancies returned by our framework.
These two plots illustrate that the formula of \cite{oketch2025} accurately captures the sign and magnitude of the true discrepancy, which is itself a small fraction of the standard error.

\begin{figure}[htbp]
\centering
\includegraphics[width=0.72\textwidth]{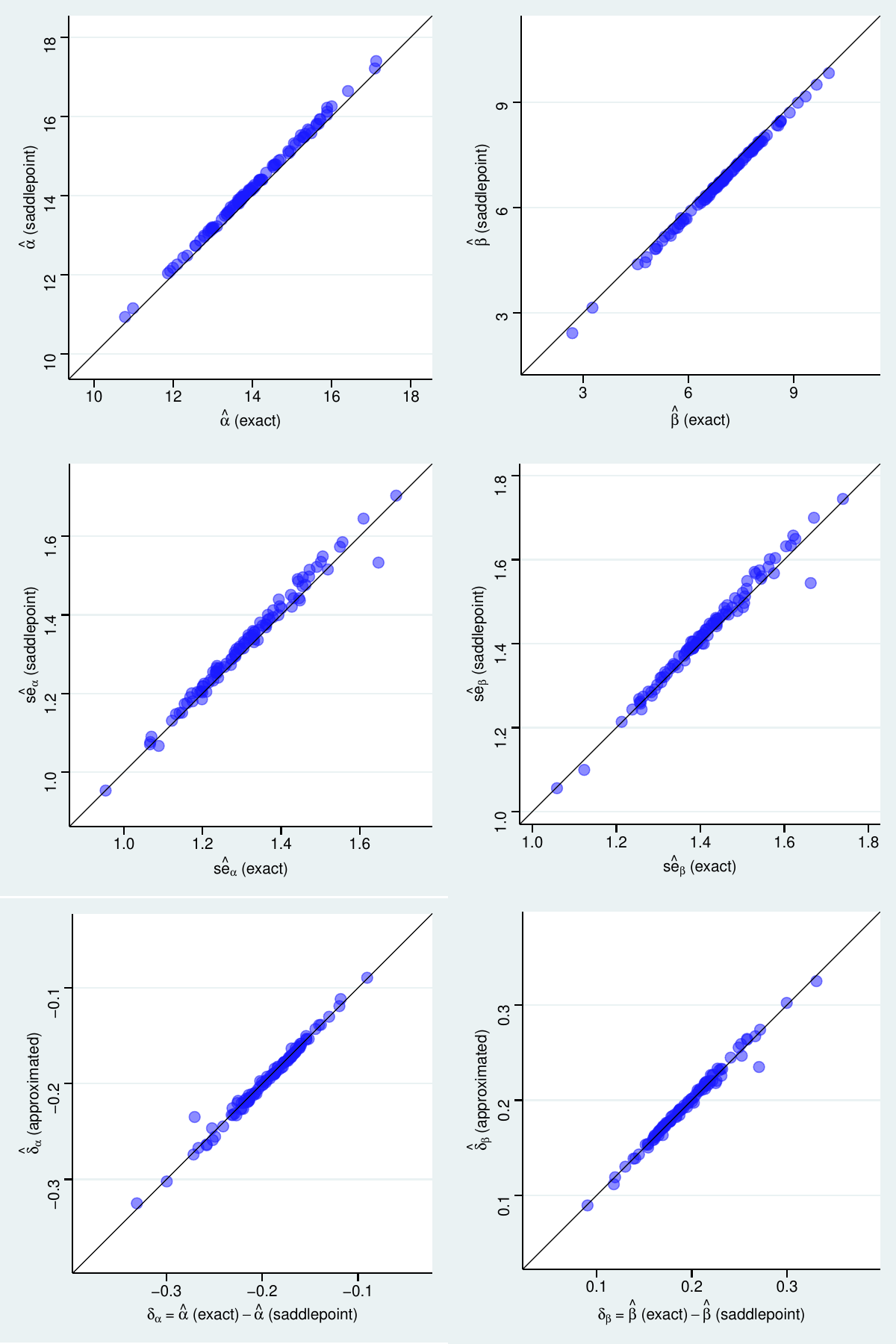}
\captionsetup{width=1.0\textwidth}
\caption{
Comparison of $\alpha$ and $\beta$ estimates (row 1), their standard errors (row 2), and the approximated versus true discrepancies (row 3),
obtained from the saddlepoint likelihood and exact likelihood for the multivariate Poisson model in \secref{sec:MVPoisson}. Each of the six panels shows estimated values based on 100 datasets simulated using $\alpha = 14$ and $\beta = 7$, with each dataset consisting of $m = 20$ i.i.d.\ copies of the vector $(Y_1, \ldots, Y_{10})$. The diagonal line is the line of equality in each panel.}
\lbfig{fig:MVPoisson}
\end{figure}

\newpage
\subsection{\texorpdfstring{Randomly stopped sum with constraints on the domain of $\boldsymbol{t}$}{Randomly stopped sum model with constraints on the domain of t}
}
\label{sec:RSS_BernoulliRV}
As a second example, suppose $U_1, \ldots, U_m$ are i.i.d.\ observations of a random variable $U$,
where
\begin{equation*}\label{eq:RSSBernoulliRV}
    U_i=\sum_{j=1}^{N_i}X_{i,j},
\end{equation*}
and where $N_1, \ldots, N_m$ are i.i.d.\ copies of an integer-valued variable $N$, and the $X_{i,j}$ are
i.i.d.\ copies of a scalar variable $X$. 
The CGF of $(U_1,\ldots,U_m)$ is 
\[
    K_{U.\mbox{\scriptsize iid}}(\boldsymbol{t})=\sum_{i=1}^m K_N(K_X(t_i)),
\] 
where $\boldsymbol{t}$ is the $m$-dimensional argument $\boldsymbol{t}=(t_1, \dots, t_m)$.

Let $X$ follow a Bernoulli distribution with parameter $p$, and let $N$ follow a Geometric distribution with known parameter $\tau$. We can show that $K_{U.\mbox{\scriptsize iid}}(\boldsymbol{t})$ is valid for 
\[
    t_i<\log\left(\frac{\tau-p\tau+p}{p(1-\tau)}\right)~\text{for}~i=1,\dots,m,
\]
where this inequality arises because the Geometric distribution exhibits a restriction on the domain of its MGF. 
Such inequality constraints are incorporated directly into our CGF framework.
Constraints for specific distributions are preprogrammed into the CGF objects we provide, so when estimating parameters using the saddlepoint likelihood, our functions will automatically account for constraints on the domain of $\boldsymbol{t}$ without any intervention from the user.

In \coderef{code:GemetricThinning} we show how to build the CGF for estimating the unknown parameter $p$. 
We use \verb|adaptor(fixed_param = tau)| to set a known value for parameter $\tau$, 
and \verb|adaptor(indices = 1)| to indicate the positions of distributional parameters in the model parameter vector. For this example, the model parameter vector consists of a single element: the Bernoulli distribution parameter $p$.

\begin{table}[htbp]
    \centering
\begin{lstlisting}[caption = CGF of the randomly stopped sum in Section \ref{sec:RSS_BernoulliRV}, label = code:GemetricThinning, linewidth=1.04\textwidth]
# Count variable CGF, Geometric(tau)
K.Ni = GeometricModelCGF(prob = adaptor(fixed_param = tau))
# Summand variable CGF, Bernoulli(p)
K.Xi = BinomialModelCGF(n = adaptor(fixed_param = 1), prob = adaptor(indices = 1))
# CGF for (U1,...,Um) where each scalar Ui is the randomly stopped sum
K.U_iid = randomlyStoppedSumCGF(count_cgf = K.Ni, summand_cgf = K.Xi, block_size = 1)
\end{lstlisting} 
\end{table}

The resulting CGF, \verb|K.U_iid|, encodes all the information necessary for maximizing the saddlepoint likelihood, including the information on the underlying inequality constraint. Hence, \verb|K.U_iid| can be passed directly as an argument to the \texttt{find.saddlepoint.MLE()} function.

In this 
example, it can be shown that the distribution of the entries simplifies to $U_i \sim \mathrm{Geometric}\left(\tau/(\tau + p-\tau p)\right)$, so $p$ can also be estimated using an exact likelihood function.
When the saddlepoint likelihood method is applied to an exponential family of distributions, the saddlepoint MLEs are identical to those obtained from exact methods \citep{Goodman2022}.
Our computation aligns with this theoretical result by returning a discrepancy value of zero, up to round-off error.

\subsection{Reusing model components}\label{sec:ReusingModelComponents}

Here we present an illustrative scenario to show how existing CGF objects can be combined, drawing on
the examples of $\boldsymbol{Y}$ and $U$ from Sections \ref{sec:MVPoisson} and \ref{sec:RSS_BernoulliRV}, respectively.

Consider an example model that features $\boldsymbol{Y}$ and $U$, where we observe
a $d$-dimensional vector $\tilde{\boldsymbol{Q}}$ given by
\begin{equation*} 
\tilde{\boldsymbol{Q}} = \sum_{j = 1}^{U} \tilde{\boldsymbol{Y}}_j,
\end{equation*} 
such that $\tilde{\boldsymbol{Q}}$ is the sum of $U$ i.i.d.\ copies of $\Tilde{\boldsymbol{Y}_j}$. Each $\Tilde{\boldsymbol{Y}_j}$ vector, corresponding to $\boldsymbol{Y}$ from \secref{sec:MVPoisson}, has $d$ entries,
while $U$ is an integer-valued scalar as defined in \secref{sec:RSS_BernoulliRV}.
This hypothetical model therefore contains three parameters denoted by $\boldsymbol{\theta} = (p, \alpha, \beta)$. 
Suppose we have previously created CGF objects \texttt{K.Y} and \texttt{K.U} using Code Blocks \ref{code:MVPoisson} and \ref{code:GemetricThinning} with the setting \texttt{m=1}.
To create the CGF object for $\tilde{\boldsymbol{Q}}$ using the previously-defined objects \texttt{K.Y} and \texttt{K.U}, we must align their notion of the model parameter with the new structure $\boldsymbol{\theta} = (p, \alpha, \beta)$. 
\coderef{code:ReusingCGFs} shows how we can reuse the CGFs in the original implementations simply by associating them with revised adapter functions for the composite model.

\begin{table}[htbp]
    \centering
\begin{lstlisting}[caption = Reusing model components in Section \ref{sec:ReusingModelComponents}, label = code:ReusingCGFs, linewidth=1.11\textwidth]
# The new parameter vector is theta = (p, alpha, beta).
# We adapt the original CGFs (K.Y and K.U) to reflect the new structure 
# of the parameter vector.
# The CGF for the summands is adapted to take the parameters 
# at indices 2 and 3 (alpha, beta).
summand.cgf = adaptCGF(cgf = K.Y, adaptor = adaptor(indices = 2:3))
# The CGF for the count is adapted to take the parameter at index 1 (p).
count.cgf = adaptCGF(cgf = K.U, adaptor = adaptor(indices = 1))
# Combine these CGFs to get the CGF of Q.
K.Q = randomlyStoppedSumCGF(count_cgf = count.cgf, summand_cgf = summand.cgf, iidReps = 1)
\end{lstlisting} 
\end{table}

\subsection{Two-source capture-recapture for estimating population size}\label{sec:Two_source_example}

For our final example, we outline the key features of a two-source capture-recapture model based on two mutually irreconcilable sampling protocols, as detailed by \citet{McClintock2013} and \citet{Bonner_Holmberg2013}, and demonstrate parameter estimation within our framework.

Consider a scenario where individually distinctive animals such as tigers are photographed from the left or right side during multiple capture occasions. 
Photographs can be matched to individual animals when they are taken from the same side (left or right), but left-side photos cannot be matched to right-side photos unless they are obtained simultaneously.

We use the following codes to specify an animal's true capture status on each capture occasion: 0 (no capture), L (captured from the left only), R (captured from the right only), B (simultaneous capture from both sides), or N (non-simultaneous capture from both sides). 
The string of capture codes across capture occasions forms
the latent identity of each animal in the population. If a latent identity contains a simultaneous capture code B, then the left and right photographs can be reconciled, but otherwise they cannot be. Identities without simultaneous captures are observed in two strands, one for the left captures and one for the right captures. For example, over four capture occasions, latent identity RLR0 is irreconcilable and is instead 
observed as two unmatched strands R0R0 and 0L00. By contrast, an identity with a simultaneous capture is directly observable, e.g., RB0L.

By comparing latent identities to observed strands, we can derive a linear transformation $\boldsymbol{Y}=\boldsymbol{AX}$, where the 
vector $\boldsymbol{Y}$ corresponds to counts of observed strands while $\boldsymbol{X}$ represents the unobserved counts of latent identities. Here, $\boldsymbol{A}$ is a known matrix that
has the effect of splintering
irreconcilable latent identities into their two observed strands. 
Although $\boldsymbol{X}$ can be modeled directly, we can only observe $\boldsymbol{Y}$, whose distribution is intractable. 
As a result, the exact likelihood function for such models cannot be computed \citep{Link2010}. 
However, using saddlepoint methods, we can approximate the probability mass function of $\boldsymbol{Y}$ and apply the saddlepoint likelihood for estimation. 

To estimate the size $N$ of the animal population, the counts $\boldsymbol{X}$ of latent identities are assumed to follow a multinomial distribution with distributional parameters $\boldsymbol{\varphi} = (N, \boldsymbol{\pi})$, where $\boldsymbol{\pi}$ is the vector of probabilities of the latent identities.
These can be modeled in terms of
the probability of obtaining left and right photo captures on any capture occasion ($p_L$ and $p_R$, respectively), and the probability of a photo capture from both sides ($p_B$), collectively representing the model parameter vector as $\boldsymbol{\theta} = (N,p_L,p_R, p_B)$. The adaptor function that relates the probability vector $\boldsymbol{\pi}$ to $\boldsymbol{\theta}$ is outlined in \secref{sec:AdaptorTwoSource} of the supplementary materials. 
Additionally, this model requires that $p_B$ be less than both $p_L$ and $p_R$ because it represents the joint probability of left and right captures.
This simultaneous constraint cannot be handled as a box constraint during optimization, highlighting the advantages of using a dedicated constrained optimization approach as the basis for our framework.

Overall, to estimate $\boldsymbol{\theta}$ for this model using the saddlepoint likelihood, we obtain the CGF of $\boldsymbol{Y}$ from the linear mapping $\boldsymbol{Y} = \boldsymbol{AX}$
and write an R function for the adaptor, $\boldsymbol{\pi} = h(\boldsymbol{\theta})$.
The adaptor function must be implemented in a form compatible with automatic differentiation software.
Finally, the constraints on the model parameters, $p_B < p_L$ and $p_B < p_R$, are incorporated 
as demonstrated in \coderef{code:Two_source_example}. The entire process of saddlepoint likelihood estimation for this complex model can therefore be carried out in a few lines of R code.
Additional details for this model and the necessary modifications to account for latent identities with zero counts are provided in \secref{sec:handling_zero_multinomialCGF} of the supplementary materials.

\begin{table}[H]
\centering
\begin{lstlisting}[caption = Pseudo-code for the capture-recapture model in \secref{sec:Two_source_example}, label = code:Two_source_example, linewidth=\textwidth]
# theta = (N, pL, pR, pB)
# Define adaptor function h with theta as the argument:
h <- function(theta) {...}

# Define the constraints on the model parameters as a  function of theta: 
# pB < pL and pB < pR
# The constraints are set to be non-positive: (pB - pL < 0, pB - pR < 0)
constraints.on.theta <- function(theta){
  list(constraints = c(theta[4] - theta[2], theta[4] - theta[3]),
       jacobian = rbind(c(0, -1, 0, 1), c(0, 0, -1, 1)))
}
# Build the CGF for Y as a function of the model parameter vector theta:
K.X <- MultinomialModelCGF(n = adaptor(indices = 1), prob.vec = h)
K.Y <- linearlyMappedCGF(cgf = K.X, matrix_A = ...)
# Find the estimate of theta:
find.saddlepoint.MLE(observed.data = Y, cgf = K.Y,
                    starting.theta = # starting.theta,
                    user.ineq.constraint.function = constraints.on.theta) 
\end{lstlisting}
\end{table}

The exact likelihood for this model is analytically intractable, so we use our discrepancy diagnostic from Section \ref{sec:discrepancy} to quantify the error introduced by maximizing the saddlepoint likelihood.
Calling \texttt{discrepancy = TRUE} in \texttt{find.saddlepoint.MLE()} yields an approximation $\hat\delta = (\hat{\delta}_N, \hat{\delta}_{p_L}, \hat{\delta}_{p_R}, \hat{\delta}_{p_B})$ that indicates how much each parameter would change if the exact likelihood were available. 
In particular, $\hat{\delta}_N$ can be read directly as the number of animals we are likely to over- or under-estimate when using the saddlepoint MLE instead of the exact MLE.
Results in \citep{oketch2025} show that $\hat{\delta}_N$ is typically an order of magnitude smaller than the sampling error of $\hat{N}$. 
In the two-source capture-recapture context, the authors have confirmed this through extensive simulations to be reported in an upcoming paper.
Thus the price we pay for using the saddlepoint approximation is negligible relative to sampling variability.

\section{Conclusion}
\label{sec:conc}

The saddlepoint approximation has recently gained traction as a tool for parameter estimation in models where exact probability density or mass functions are computationally intractable.
Despite the clear advantages of the method, uptake remains limited.
This work introduces a computational framework, implemented in the \verb|saddlepoint| R package, that simplifies the application of saddlepoint methods and streamlines the estimation process. 
Our aim is to make saddlepoint methods more accessible, promoting their wider adoption.

Our framework allows modelers to construct CGFs using the building blocks outlined in Section \ref{sec:ModelBuildingBlocks}, enabling them to concentrate exclusively on 
the structure of their model and side-stepping the intricate details often associated with saddlepoint methods. By employing constrained maximization of the saddlepoint likelihood
and leveraging automatic differentiation tools for gradient-based optimization,
we ensure that our framework is both computationally robust and capable of accommodating any model-specific constraints. 
Practitioners can also incorporate custom CGF objects and operations into the software.

Beyond the high-level functions \texttt{find.saddlepoint.MLE()} and \texttt{compute.spa.negll()}, the package provides a range of lower-level utilities; see the package documentation for examples and additional functionality.
More broadly, the framework provides a convenient interface for manipulating MGFs and CGFs, thereby supporting a wide range of statistical applications beyond parameter estimation.

\section{Supplementary material}

The file \textit{Supplement to `A general framework for computation and estimation using the saddlepoint approximation'} contains supplementary material for the article, including expanded descriptions of the algorithms, additional notes for each modeling example, and guidance on specifying user-defined constraints.
A separate file, saddlepoint\_code.zip, contains the complete, ready-to-run R scripts for each code block referenced in the paper, together with a README file and additional instructions.

\section{Disclosure statement}\label{disclosure-statement}

The authors report there are no competing interests to declare.

\section{Declaration of generative AI use}

The authors report generative AI was not used in their research or in the preparation of this manuscript.

\section{Funding}

The authors gratefully acknowledge funding support from the Royal Society of New Zealand Marsden fund.

\bibliography{bibliography_file.bib}

\newpage

\setcounter{section}{0}
\renewcommand{\thesection}{S\arabic{section}}
\renewcommand{\theHsection}{S\arabic{section}}
\setcounter{equation}{0}
\renewcommand{\theequation}{S\arabic{equation}}
\renewcommand{\theHequation}{S\arabic{equation}}
\setcounter{table}{0}
\renewcommand{\thetable}{S\arabic{table}}
\renewcommand{\theHtable}{S\arabic{table}}
\setcounter{figure}{0}
\renewcommand{\thefigure}{S\arabic{figure}}
\renewcommand{\theHfigure}{S\arabic{figure}}

\makeatletter
\renewcommand\section{\@startsection{section}{1}{\z@}%
  {-3.5ex \@plus -1ex \@minus -.2ex}%
  {2.3ex \@plus.2ex}%
  {\normalfont\large\bfseries}}
\makeatother

\providecommand{\Vec}{\vec}

\begin{center}
  {\Large\bf Supplement to ``A general framework for computation and estimation using the saddlepoint approximation''}
\end{center}
\bigskip

\section{Remarks on the two-step approach to maximization of the saddlepoint likelihood}
\label{sec:two_step_approach}

Given a $d$-dimensional, $\boldsymbol{\theta}$-dependent random variable $\boldsymbol{Y}$, the saddlepoint log-likelihood function is
\begin{equation}\label{eqSuppl:SaddlepointLogLik}
    \begin{aligned}
          & \log \hat{L} (\boldsymbol{\theta}\,|\,\boldsymbol{y})
          =
          K_Y(\hat{\boldsymbol{t}};\boldsymbol{\theta}) - \hat{\boldsymbol{t}}\boldsymbol{y} - \tfrac{d}{2} \log(2\pi) - \tfrac{1}{2} \log \det ( K_Y''(\hat{\boldsymbol{t}};\boldsymbol{\theta}) ),
    \end{aligned}
\end{equation}
where $\hat{\boldsymbol{t}} = \hat{\boldsymbol{t}}(\boldsymbol{\theta};\boldsymbol{y})$
is a $d$-dimensional row vector that
solves the saddlepoint equation $K_Y'(\hat{\boldsymbol{t}};\boldsymbol{\theta})=\boldsymbol{y}$.
Most applications of saddlepoint likelihood estimation to date have employed a nested optimization strategy to maximize \eqref{eqSuppl:SaddlepointLogLik}. The following two-step procedure summarizes this approach.

\begin{enumerate}[label=(\arabic*)]
    \item {} [Inner optimization] For a specified value of $\boldsymbol{\theta}$, find the value of $\boldsymbol{t}$ that minimizes $K_Y(\boldsymbol{t};\boldsymbol{\theta}) - \boldsymbol{ty}$. This serves as an expression of the saddlepoint equation, i.e.,
    \begin{equation}\label{eq:argmin_t_hat}
        \hat{\boldsymbol{t}} = \underset{\boldsymbol{t}}{\mathrm{argmin}}\, K_Y(\boldsymbol{t};\boldsymbol{\theta}) - \boldsymbol{ty}.
    \end{equation}
    \item {} [Main/outer optimization] Find the value of $\boldsymbol{\theta}$ that maximizes \eqref{eqSuppl:SaddlepointLogLik}.
    The evaluation of \eqref{eqSuppl:SaddlepointLogLik} at each $\boldsymbol{\theta}$ involves an iteration of the inner optimization to find the corresponding $\hat{\boldsymbol{t}}$ for that $\boldsymbol{\theta}$.
\end{enumerate}
Here we offer some remarks about the two-step procedure. These considerations motivated our decision to use an alternative method of constrained optimization as the default for our computational framework.
\begin{itemize}
    \item The successful update of $\boldsymbol{\theta}$ in step 2 depends on the accuracy of the current value of $\hat{\boldsymbol{t}}$.
    In other words, the outer loop's convergence hinges on the inner loop's precision.

    \item Updating $\hat{\boldsymbol{t}}$ in the inner loop utilizes the current value of $\boldsymbol{\theta}$. At initial iterative stages, this $\boldsymbol{\theta}$ may differ significantly from the
    final optimal value of $\boldsymbol{\theta}$.
    Committing to an accurate computation of $\hat{\boldsymbol{t}}$, with a notably incorrect $\boldsymbol{\theta}$, especially in initial iterations,
    can be an inefficient use of computational time.

    \item For the maximization in step 2, the gradient computation is intricate because it involves the gradient of the inner optimization problem.
This gradient can be derived manually or computed via automatic differentiation \citep{TMB2016}.
However, this gradient computation and the problem as a whole become more intricate when the CGF $K_Y(\boldsymbol{t};\boldsymbol{\theta})$ is only valid for constrained values of $\boldsymbol{t}$.
The constrained approach discussed in \secref{sec:ConstrainedOptim} of the main text circumvents these complexities by embodying constraints on $\boldsymbol{t}$ into the constrained objective function, along with the constraint implied by the saddlepoint equation, and any other constraints on the model parameters $\boldsymbol{\theta}$.

\end{itemize}

\section{Adaptor function for the two-source capture recapture model}
\label{sec:AdaptorTwoSource}

For the two-source capture-recapture model described in \secref{sec:Two_source_example}
of the main text,
the model parameters are the probabilities of any left-side capture $(p_L)$, any right-side capture $(p_R)$, and any simultaneous left and right captures $(p_B)$ on a single capture occasion.
Based on these parameters, we need to assign probabilities to each of the capture statuses 0, L, R, B, and N outlined in \secref{sec:Two_source_example}.
To derive these probabilities, we consider
that captures occur according to an underlying encounter process, where the
number of encounters with an animal on a single capture occasion is $E \sim \Poisson(\mu)$.
Define the following probabilities for each encounter,
\begin{equation*}
    \begin{aligned}
        & \Pr(\text{animal is captured from the left}) = \alpha, \\
        & \Pr(\text{animal is captured from the right}) = \beta.
    \end{aligned}
\end{equation*}
Assuming independence of left-right captures within each encounter, then for a single occasion we marginalize over the number of encounters, $E$, giving
\begin{itemize}
    \item $p_L = \Pr(\text{any left side capture}) = 1 - \exp(-\mu\alpha)$,
    \item $p_R = \Pr(\text{any right side capture}) = 1 - \exp(-\mu\beta)$,
    \item $p_B = \Pr(\text{any simultaneous left and right captures}) = 1 - \exp(-\mu\alpha\beta)$.
\end{itemize}
Based on these assumptions, we obtain a parametric form for all five probabilities $\Pr(\mathrm{0})$, $\Pr(\mathrm{L}), \Pr(\mathrm{R})$, $\Pr(\mathrm{B})$, and $\Pr(\mathrm{N})$ in terms of the three model probabilities $p_L$, $p_R$, and $p_B$. After some algebra, for each capture occasion, we have
\begin{equation}\label{eq:Probability_of_capture_status}
    \begin{aligned}
        & \Pr(0) = (1-p_L)(1-p_R)/(1-p_B), \\
        & \Pr(\mathrm{L}) = 1 - p_R - \Pr(0), \\
        & \Pr(\mathrm{R}) = 1 - p_L - \Pr(0), \\
        & \Pr(\mathrm{B}) = p_B, \\
        & \Pr(\mathrm{N}) = p_L + p_R - p_B - 1 + \Pr(0).
    \end{aligned}
\end{equation}

Denote the set of all possible latent identities by $H$.
For a latent identity $\lambda \in H$, the probability $\pi_\lambda$ is the product of the probabilities in \eqref{eq:Probability_of_capture_status} over all capture occasions.
For instance
over four occasions, one of the latent identities is $\lambda = \mathrm{RLR0}$ with probability
$\pi_\lambda =  \Pr(\text{R})\Pr(\text{L})\Pr(\text{R})\Pr(\text{0})$.
Finally, in the formulation $\boldsymbol{Y}=\boldsymbol{AX}$ from  \secref{sec:Two_source_example} of the main text, the observed counts $\boldsymbol{Y}$ with model parameter $\boldsymbol{\theta}=(N, p_L, p_R, p_B)$ are described by specifying the latent counts $\boldsymbol{X}$ in terms of their multinomial parameter $\boldsymbol{\phi}=(N,\boldsymbol{\pi})$. The adaptor function links $\boldsymbol{\theta}$ to $\boldsymbol{\phi}$ by computing $\pi_\lambda$ for all $\lambda\in H$. See the accompanying code for the full implementation.

\section{Handling a zero-outcome category in the multinomial CGF}
\label{sec:handling_zero_multinomialCGF}

Consider the random vector $\boldsymbol{Y} = (Y_1,\dots,Y_d, Y_{d+1})$ which follows a multinomial distribution with parameters $N$ and the probability vector $\boldsymbol{\pi} = (\pi_1,\dots,\pi_d, \pi_{d+1})$, where $\sum_{i=1}^{d+1} \pi_i = 1$.
Then
\begin{equation}\label{eqDisc:MultinomialCGF}
    K_Y(\boldsymbol{t}; N, \boldsymbol{\pi}) = N\log \left(\sum_{i=1}^{d+1} \pi_i \exp (t_i)\right),~\boldsymbol{t} \in \mathbb{R}^{d+1}.
\end{equation}
If we condition on one of the counts in $\boldsymbol{Y}$ being zero, say $\boldsymbol{W} \sim (\boldsymbol{Y}\, | \, Y_{d+1}=0)$, then $\boldsymbol{W}$ also has a multinomial distribution with parameters
$N$ and the probability vector $\Vec{\boldsymbol{p}} = (p_1,\dots,p_d)$, where $p_i = \pi_i (\sum_{j=1}^d \pi_j)^{-1}$.

If instead we restrict to the event $\{Y_{d+1}=0\}$, the corresponding restricted CGF is
defined to include only those terms of the expectation involving $Y_{d+1}=0$, and is
equivalent to setting
$t_{d+1} = -\infty$ in the CGF of $\boldsymbol{Y}$.
We introduce the notation $K_{\mathrm{subunitary}}$ to represent this restricted CGF, which we define as follows:
\begin{equation*}
    \begin{aligned}
        &K_{\mathrm{subunitary}}(t_1,\dots,t_d; N, \pi_1,\dots, \pi_d)
        \\
        &\qquad\qquad\qquad\qquad
        =
        \log \mathbb{E}\left\{ \exp(t_1 Y_1 + \dots + t_d Y_d) \mathbbm{1}(Y_{d+1}=0) \right\},
        \\
        &\qquad\qquad\qquad\qquad
        =
        K_Y(t_1, \dots, t_d, -\infty; N, \pi_1, \dots, \pi_d, 1 - \textstyle\sum_{i=1}^d \pi_i),
        \\
        &\qquad\qquad\qquad\qquad
        =
        N\log \left( \sum_{i=1}^{d} \pi_i \exp(t_i) \right),
        \\
        &\qquad\qquad\qquad\qquad
        =
        N\log \left( \sum_{i=1}^{d} \pi_i \right) +
        N\log
        \left\{
            \sum_{i=1}^{d}
            \left(
            \frac{\pi_i}{\sum_{j=1}^{d} \pi_j}
            \right)
            \exp(t_i)
        \right\},
        \\
        &\qquad\qquad\qquad\qquad
        =
        N \log \left( \sum_{i=1}^{d} \pi_i \right) +
        N \log \left( \sum_{i=1}^{d} p_i \exp(t_i) \right),
        \\
        &\qquad\qquad\qquad\qquad
        =
        \log\{\P(Y_{d+1}=0)\} +
        K_{W}(t_1,\dots,t_d; N, \Vec{\boldsymbol{p}}),
    \end{aligned}
\end{equation*}
where $\boldsymbol{W} \sim (\boldsymbol{Y} | Y_{d+1}=0)$ was defined above.
From this, and equation~\eqref{eq:Saddlepoint_Defn} in the main text, it can be seen that the saddlepoint approximation using the subunitary CGF, $K_{\mathrm{subunitary}}$, is equivalent to
multiplying the saddlepoint approximation using the CGF of $\boldsymbol{W}$, $K_{W}$, by the factor $\P(Y_{d+1}=0)$.
The term `subunitary' indicates that the probability vector $\Vec{\boldsymbol{p}}$ sums to less than one.

The scenario outlined here is relevant in capture-recapture models, where certain latent identities are known to have zero counts because they are incompatible with any observed capture strands. In such models, the sum of the probabilities associated with all other latent identities is less than 1.
We use this subunitary CGF to implement the saddlepoint likelihood approach in multinomial models with known zero counts, where zero counts are merged into a single cell represented by cell $d+1$ above.
If we do not use the subunitary CGF for zero counts, the corresponding saddlepoint equation
$K_Y'(\hat{\boldsymbol{t}}) = \boldsymbol{y}$ has no finite solution.
Even if we fix the non-finite component at $\hat{t}_{d+1} = -\infty$ instead of including it in the optimization, the determinant of the matrix $K_Y''(\hat{\boldsymbol{t}}; \boldsymbol{\theta})$ (see Equation~\eqref{eq:Saddlepoint_Defn} in the main text) is zero, rendering the saddlepoint approximation undefined. Using the subunitary CGF avoids these problems.

This modification of the multinomial CGF is incorporated into our framework and is implemented through the R function
\[    \verb|SubunitaryMultinomialModelCGF(n = ..., prob_vec = ...)|.
\]
See the accompanying code for an implementation of this CGF in the two-source model in \secref{sec:Two_source_example}.

\section{Adding an external CGF into the framework}
\label{CustomCGF}

Here we provide a template for practitioners to create new CGF objects within our framework. We demonstrate its use with a zero-truncated Poisson random variable,
using the aggregate claims model of
\cite{Meraou2022} described in \secref{sec:RSS_section} of the main text as an example setting.
Note that the zero-truncated Poisson distribution is simply used here as an example to showcase the extensibility of our framework.

Consider the total insurance payout defined as
\begin{equation}\label{eqDisc:ZeroTruncPoisExp}
Y = \sum_{i = 1}^{L} T_i,
\end{equation}
where the unobserved payouts for single claims are modeled by $T_i \sim \mathrm{Exponential}(\beta)$ and the unobserved number of approved claims, $L$, follows
a zero-truncated Poisson distribution with parameter $\theta$.
This expression represents a randomly stopped summation of i.i.d.\ exponentially distributed random variables.

To estimate the parameter vector $\boldsymbol{\eta} = (\theta, \beta)$ via the saddlepoint likelihood method, we note that
the CGF for the exponential distribution and the randomly stopped sum operation corresponding to \eqref{eqDisc:ZeroTruncPoisExp} are preprogrammed into our R package. However, the CGF of a zero-truncated Poisson is not provided, so this component must be added.

In our software framework, we add new CGF objects using the constructor shown below. Its required arguments are functions that compute
the vectorized CGF and its first four derivatives with respect to $\boldsymbol{t}$.
The CGF object is constructed as follows:
\begin{verbatim}
                createCGFfromVectorizedFunctions(
                        K_vectorized  = ...,
                        K1_vectorized = ...,
                        K2_vectorized = ...,
                        K3_vectorized = ...,
                        K4_vectorized = ...
                )
\end{verbatim}
For additional examples, see \url{https://godrick.github.io/saddlepoint/articles/custom-cgf.html}.

The CGF for $m$ i.i.d.\ copies of a zero-truncated Poisson random variable is shown below for $\boldsymbol{t} \in \mathbb{R}^m$:
\begin{equation}
    K_{L}(\boldsymbol{t};\theta) = \sum_{i = 1}^m \left\{ -\theta - \log\left(1 - e^{-\theta}\right) + \log(e^{\theta e^{t_i}} - 1) \right\}.
\end{equation}
Component $i$ of the gradient \(K'_{L}(\boldsymbol{t};\theta)\) is given by
\begin{equation}
    \frac{ \partial K_L (\boldsymbol{t}; \theta) } {\partial t_i}
    =
    \frac{\theta e^{\theta e^{t_i} + t_i}}{e^{\theta e^{t_i}} - 1}.
\end{equation}
Elements of the diagonal matrix \(K''_{L}(\boldsymbol{t};\theta)\) are
\begin{equation}
    \frac{ \partial^2 K_L (\boldsymbol{t}; \theta) } {\partial t_i^2}
    =
    \frac{ \partial K_L (\boldsymbol{t}; \theta) } {\partial t_i}
    -
    \left(\frac{ \partial K_L (\boldsymbol{t}; \theta) } {\partial t_i}\right)^2
    +
    \frac{\theta^2 e^{2t_i + \theta e^{t_i}}}{e^{\theta e^{t_i}} - 1} .
\end{equation}
The required third and fourth derivatives may be obtained analytically. In the accompanying code, they are obtained using the R package \texttt{Deriv}.
After supplying these five functions to \verb|createCGFfromVectorizedFunctions()|, the resulting CGF object can be used for model construction and saddlepoint likelihood estimation, as illustrated in the following Code Block.

\begin{table}[htbp]
    \centering
\begin{lstlisting}[caption = CGF of the Zero-Truncated Poisson-Exponential model, label = code:ZeroTruncated, linewidth=1.01\textwidth]
library(saddlepoint)
# Create the custom CGF object
ZeroTruncatedPoissonCGF <- createCGFfromVectorizedFunctions(...)

# Adapt the ZeroTruncatedPoissonCGF to create a CGF for L
# that acknowledges parameter dependency (theta, beta)
L.CGF <- adaptCGF(cgf = ZeroTruncatedPoissonCGF, adaptor = adaptor(indices = 1))

# Build CGF for Exponential distribution
T.CGF <- ExponentialModelCGF(rate = adaptor(indices = 2))

# Build the CGF for Y
K.Y = randomlyStoppedSumCGF(count_cgf = L.CGF, summand_cgf = T.CGF, block_size = 1)

# Estimate MLEs using saddlepoint likelihood
MLEs = find.saddlepoint.MLE(observed.data = Y,
                               cgf = K.Y,
                               starting.theta = c(0.5, 0.5),
                               lb.theta = c(0.01,0.01),
                               std.error = TRUE)
\end{lstlisting}
\end{table}

\end{document}